\documentclass[apj,numberedappendix,appendixfloats,twocolappendix]{emulateapj}

\usepackage{natbib}
\usepackage{amsmath}
\usepackage{xcolor}
\usepackage[english]{babel}
\usepackage{graphicx}
\usepackage{xspace}
\usepackage{enumitem}

\newcommand{\hMpc}{{\ifmmode{h^{-1}{\rm Mpc}}\else{$h^{-1}$Mpc}\fi}}
\newcommand{\Mpc}{{\ifmmode{{\rm Mpc}}\else{Mpc}\fi}}
\newcommand{\hkpc}{{\ifmmode{h^{-1}{\rm kpc}}\else{$h^{-1}$kpc}\fi}}
\newcommand{\kpc}{{\ifmmode{ {\rm kpc} }\else{{\rm kpc}}\fi}}
\newcommand{\kms}{{\ifmmode{ {\rm km\,s^{-1}} }\else{ ${\rm km\,s^{-1}}$ }\fi}}
\newcommand{\hMsun}{{\ifmmode{h^{-1}{\rm {M_{\odot}}}}\else{$h^{-1}{\rm{M_{\odot}}}$}\fi}}
\newcommand{\Msun}{{\ifmmode{{\rm M}_{\odot}}\else{${\rm M}_{\odot}$}\fi}}
\newcommand{\Mhalo}{{\ifmmode{M_{\rm halo}}\else{$M_{\rm halo}$}\fi}}
\newcommand{\Rvir}{{\ifmmode{R_{\rm vir}}\else{$R_{\rm vir}$}\fi}}
\newcommand{\Mvir}{{\ifmmode{M_{\rm vir}}\else{$M_{\rm vir}$}\fi}}
\newcommand{\Mstar}{{\ifmmode{M_{\rm star}}\else{$M_{\rm star}$}\fi}}
\newcommand{\Vrot}{{\ifmmode{V_{\rm rot}}\else{$V_{\rm rot}$}\fi}}
\newcommand{\ltsima}{$\; \buildrel < \over \sim \;$}
\newcommand{\gtsima}{$\; \buildrel > \over \sim \;$}
\newcommand{\lsim}{\lower.5ex\hbox{\ltsima}}
\newcommand{\gsim}{\lower.5ex\hbox{\gtsima}}

\def\lesssim{\mathrel{\hbox{\rlap{\hbox{\lower4pt\hbox{$\sim$}}}\hbox{$<$}}}}
\def\gtrsim{\mathrel{\hbox{\rlap{\hbox{\lower4pt\hbox{$\sim$}}}\hbox{$>$}}}}

\newcommand{\beq}{\begin{equation}}
\newcommand{\eeq}{\end{equation}}
\def\beqa{\begin{eqnarray}}
\def\eeqa{\end{eqnarray}}
\def\LCDM{\ensuremath{\Lambda}CDM}

\def\head{ \vbox to 0pt{\vss \hbox to 0pt{\hskip 440pt\rm
      LA-UR-10-07069\hss} \vskip 25pt}}

\def \kms {\ifmmode  \,\rm km\,s^{-1} \else $\,\rm km\,s^{-1}  $ \fi }
\def \kpc {\ifmmode  {\rm kpc}  \else ${\rm  kpc}$ \fi  }  
\def \hkpc {\ifmmode  {h^{-1}\rm kpc}  \else ${h^{-1}\rm kpc}$ \fi  }  
\def \hMpc {\ifmmode  {h^{-1}\rm Mpc}  \else ${h^{-1}\rm Mpc}$ \fi  }  
\def \Mpch {\ifmmode  {h^{-1}\rm Mpc}  \else ${h^{-1}\rm Mpc}$ \fi  }  
\def \Msun {\ifmmode {\rm M}_{\odot} \else ${\rm M}_{\odot}$ \fi} 
\def \hMsun {\ifmmode h^{-1}\,\rm M_{\odot} \else $h^{-1}\,\rm M_{\odot}$ \fi}

\def \LCDM {\ifmmode \Lambda{\rm CDM} \else $\Lambda{\rm CDM}$ \fi}
\def \sig8 {\ifmmode \sigma_8 \else $\sigma_8$ \fi} 
\def \OmegaM {\ifmmode \Omega_{\rm m} \else $\Omega_{\rm m}$ \fi} 
\def \Omegab {\ifmmode \Omega_{\rm b} \else $\Omega_{\rm b}$ \fi} 
\def \OmegaL {\ifmmode \Omega_{\rm \Lambda} \else $\Omega_{\rm \Lambda}$\fi} 
\def \Deltavir {\ifmmode \Delta_{\rm vir} \else $\Delta_{\rm vir}$ \fi}
\def \rhocrit {\ifmmode \rho_{\rm crit} \else $\rho_{\rm crit}$ \fi}
\def \rhou {\ifmmode \rho_{\rm u} \else $\rho_{\rm u}$ \fi}
\def \zc {\ifmmode z_{\rm c} \else $z_{\rm c}$ \fi}

\def\lcdm{\ensuremath{\Lambda\textrm{CDM}}\xspace}

\def\lcdm{\ensuremath{\Lambda\textrm{CDM}}\xspace}

\slugcomment{submitted to ApJ}
\shorttitle{The Milky Way's central stellar populations}
\shortauthors{Buck, Ness, Macci\`o, et al.}

\begin{document}

\title{Stars behind bars I: The Milky Way's central stellar populations} 

\author{Tobias Buck\altaffilmark{$\star$}\altaffilmark{1}\altaffilmark{$\dagger$}, Melissa K. Ness\altaffilmark{1}, Andrea V. Macci\`o\altaffilmark{2,1}, Aura Obreja\altaffilmark{3}, Aaron A. Dutton\altaffilmark{2}}

\altaffiltext{$\star$}{buck@mpia.de}
\altaffiltext{1}{Max-Planck Institut f\"ur Astronomie, K\"onigstuhl 17, D-69117 Heidelberg, Germany}
\altaffiltext{2}{New York University Abu Dhabi, PO Box 129188, Abu Dhabi, UAE}
\altaffiltext{3}{Universit\"ats-Sternwarte M\"unchen, Scheinerstraße 1, D-81679 M\"unchen, Germany}
\altaffiltext{$\dagger$}{Member of the International Max Planck Research School for Astronomy and Cosmic Physics at the University of Heidelberg, IMPRS-HD, Germany.}

\begin{abstract}
We show for the first time, that a fully cosmological hydrodynamical simulation can reproduce key properties of the innermost region of the Milky Way (MW). Our high resolution simulation reproduces qualitatively the profile and kinematics of the MW's boxy/peanut-shaped bulge, and hence we can use it to reconstruct and understand the bulge assembly. In particular, the age dependence of the X-shape morphology of the simulated bulge parallels the observed metallicity dependent split in the red clump stars of the inner Galaxy. 
We use this feature to propose an observational metric that (after calibrated against a larger set of simulations) might allow us to quantify when the bulge formed from the disk. The metric we propose can be employed with upcoming survey data to constrain the age of the MW bar. From the split in stellar counts we estimate the formation of the 4~kpc scale bar in the simulation to have happened  $t^{\rm bar}_{\rm form}\sim8^{+2}_{-2}$ Gyr ago, in good agreement with conventional methods to measure bar formation in simulations. We test the prospects for observationally differentiating the stars that belong to the bulge/bar compared to the surrounding disk, and find that the inner disk and bulge are practically indistinguishable in both chemistry and ages. 

\end{abstract}

\section{Introduction} \label{sec:introduction}

Almost 50 percent of all nearby galaxies show signs of a boxy or peanut shaped bulge \citep{Lutticke2004} and our galaxy, the Milky Way (hereafter, MW), is observed to host a boxy/peanut shaped bulge and a galactic bar \citep{Okuda1977,Blitz1991,Weiland1994,Dwek1995}. The major axis of the MW's Galactic bar is inclined by about $27^\circ$ with respect to the line-of-sight and it reaches out to about $3.5$ kpc \citep{Gerhard2002, Wegg2015} with the bar extending in the plane up to about 5 kpc \citep{Portail2017}.

The formation scenario of boxy/peanut-shaped bulges from the galactic disk is well studied in idealized simulations and several mechanisms have been identified where by stars of the disk become the boxy bulge. Isolated $N$-body simulations of galaxies have shown that boxy/peanut bulges can form in-situ via disc instabilities \citep{Bureau1999,Athanassoula2009}, where flat discs develop a bar after only a few revolutions. This bar then puffs up into a boxy/peanut bulge structure \citep{Raha1991,Merritt1994,Bureau2005, Debattista2006} via a vertical instability, the so called buckling instability. The formation of boxy/peanut bulges has also been explained by orbit trapping, into a vertical Lindblad resonance during bar growth \citep{Combes1981,Quillen2002,Quillen2014} and via orbits associated with vertical resonances \citep{Combes1990,Pfenniger1991}. However, there is no overall agreement as to what  specific orbits and in what relative fraction actually make up the boxy/peanut bulge of the Galactic bulge of the MW \citep[see e.g.][]{Portail2015a,Portail2015}.

The observational evidence suggests that the MW's boxy/peanut bulge has, in large part, formed from the disk \citep[e.g.][]{Ness2012}. However, the time of formation and the details of the subsequent evolution are uncertain. Additionally, the fraction of any underlying component that is not associated with the disk and whether this is a classical bulge or part of the inner halo, is under debate.  

The bulge is not only observed to be boxy in photometric images, but the red clump stars in the center of our Galaxy are split into two components well separated along the line of sight \citep{Nataf2010,McWilliam2010}. The interpretation of this phenomenon is an underlying X-shaped structure in the bulge \citep{Li2012,Saito2011, Ness2012}, a clear signature of formation from the disk. This split shows different properties for different metallicity populations; the metal rich populations show the strongest split, the metal poor stars a weaker split, with no split seen below [Fe/H] $<$ -0.5 \citep[e.g.][]{Ness2012,Uttenthaler2012,Rojas-Arriagada2014}.

Many $N$-body models of bulge formation starting from a pure thin disc with evolving disc instabilities can alone explain the \textit{overall} observed characteristics of the MW's bulge \citep{Martinez2013, Vasquez2013,Gardner2014,Zoccali2014}. Data from the BRAVA \citep{Howard2008}, ARGOS \citep{Freeman2013} and APOGEE surveys \citep{Apogee} have revealed cylindrical rotation in the bulge of the MW \citep{Howard2009,Ness2013a,Ness2016a} which is characteristic of an in-situ boxy/peanut bulge. Moreover, using BRAVA kinematics \citep{Kunder2012}, \cite{Shen2010Bulge} constrained any merger-generated component of the bulge to be less than 8 percent. 

However, the stars in the MW's bulge show not only different morphology as a function of metallicity \citep[][e.g.]{Dekany2013,Portail2016} but also different kinematics. This is most dramatic for the small fraction ($\sim$ 5\%) of most metal poor stars in the bulge, [Fe/H] $\le$ -1 dex \citep{Kunder2016, Ness2013}, but is seen across the full metallicity distribution function, which extends from --3 $<$ [Fe/H] $<$ 0.6 dex. These differences have been explained as being a consequence of composite populations in the inner MW.  Several studies \citep[e.g.][]{Clarkson2008,Hill2011,Gonzalez2011} have decomposed the MDF of the bulge into two stellar populations, which have different kinematics. These authors associate the metal-rich, rapidly rotating and dynamically cold population they find with a boxy/peanut bulge with disc origin. The metal-poor component, dynamically hotter, more slowly rotating population has been associated to be an old classical bulge component  \citep[e.g.][]{Rojas-Arriagada2017} although no final agreement about this interpretation exists \citep[e.g.][]{Zoccali2016}.

Using the ARGOS survey, \cite{Ness2013} showed that the metallicity distributions of stars in the inner Galaxy can be represented by five different components, which these authors associated with the thin disk, the thick disk, the boxy peanut bulge, and also a very small population at the lowest metallicities, of metal weak thick disk and halo.  \citet{Debattista2016} has explained these composite populations, where by the stars have different properties as a function of metallicity, using idealized $N$-body simulations and isolated hydrodynamical simulations of galaxy formation, largely in initial radial velocity dispersion, which leads to separation in evolution. They show that a bar is able to separate initially co-spatial populations of stars if only their initial radial velocity dispersion is different.  \citet{Portail2017} have used these observational data to constrain the density distribution and morphology of stars in the inner region with their associated properties and find distinct separation in morphology are associated with the different observed kinematics as a function of metallicity. \citet{DiMatteo2015} use the observational data with their N-body simulation to determine that both thin and thick disk are necessary to explain the observed kinematic behavior in the MW.

While isolated simulations and controlled $N$-body experiments \citep[e.g.][]{DiMatteo2016,Athanassoula2017,Fragkoudi2017}  are ideal to study the mechanisms at play in shaping the bulge and are able to well explain the observed kinematics, they ultimately exclude the chemical enrichment history of star formation and are not able to properly model different temporal or chemical populations. Isolated hydrodynamical simulations of galaxy formation, which include star formation and feedback do alleviate the problem of self-consistent chemical enrichment and can well explain (compare) metallicity trends. Nonetheless, even these simulations miss one key ingredient to study the build-up of the bulge in a realistic environment. Galaxies grow in a cosmological environment by accreting gas, forming stars, getting disturbed and bombarded with satellite galaxies and self enrich the gas with metals by stellar feedback. Including all these phenomena is absolutely necessary in order to study the different components of galactic bulges and explain their origin. Furthermore we need to consistently be able to reproduce properties of the universe on the smallest and largest scales.

Cosmological simulations are the means to link and understand the far and near universe, but have typically  had a hard time reproducing realistic bulges observed in spiral galaxies. Due to its inherent hierarchical nature, \lcdm simulations of galaxy formation predict that galactic spheroids are primarily built up through hierarchical mergers \citep[e.g.][]{Kauffmann1993,Abadi2003,Kobayashi2011,Guedes2013} which produce an old classical bulge, incompatible with a boxy/peanut shaped bulge, as is observed in our Galaxy \citep{Weiland1994,Dwek1995}.

In this paper we use, for the first time, a fully cosmological hydrodynamical simulation of galaxy formation to study the inner region of a galaxy showing a bulge that is like that of the MW.  We will establish the similarity of the simulated bulge with the observed features of the MW bulge  - and critically - we can make predictions for the properties of the bulge, the bar and the surrounding disk in age, chemistry and dynamics. By comparing the properties of bulge stars to properties of other components of this galaxy (thin, thick disc or halo) and by tracing the stars from the inner region through time we can understand the mass assembly of the bulge and disentangle effects of secular evolution from accretion.

This paper is organized as follows: In \S2 we present the model galaxy, its general properties and describe the simulation code. We continue in \S3 with a comparison of general bulge properties of our simulation and properties including the rotation and dispersion profiles of bulge stars and the split in the line-of-sight counts of stars, which we examine in different age bins. We develop an observational metric to measure the age of the bar from the split, a critical step toward making quantifications from simulations to directly test with new generations of surveys.  In \S4 we turn to compare properties of stars in the bar with stars in the disc showing neither chemistry nor ages will allow the populations to be distinguished. Finally in \S5 we summarize our results and conclude.

\begin{table}
\label{tab:sims}
\begin{center}
\caption{Simulation properties.}
\begin{tabular}{c c c c}
		\hline\hline
		property & particle mass & Force soft. & Smoothing length\\
		  & [$10^5$ \Msun] & [pc] & (median, min.) [pc] \\
		\hline
		DM & 5.141 & 620 & - \\
		GAS & 0.938 & 265 & (155, 20) \\
		STARS & 0.313 & 265 & - \\
		\hline
\end{tabular}
\end{center}
\end{table}

\begin{figure}
\includegraphics[width=1.075\columnwidth]{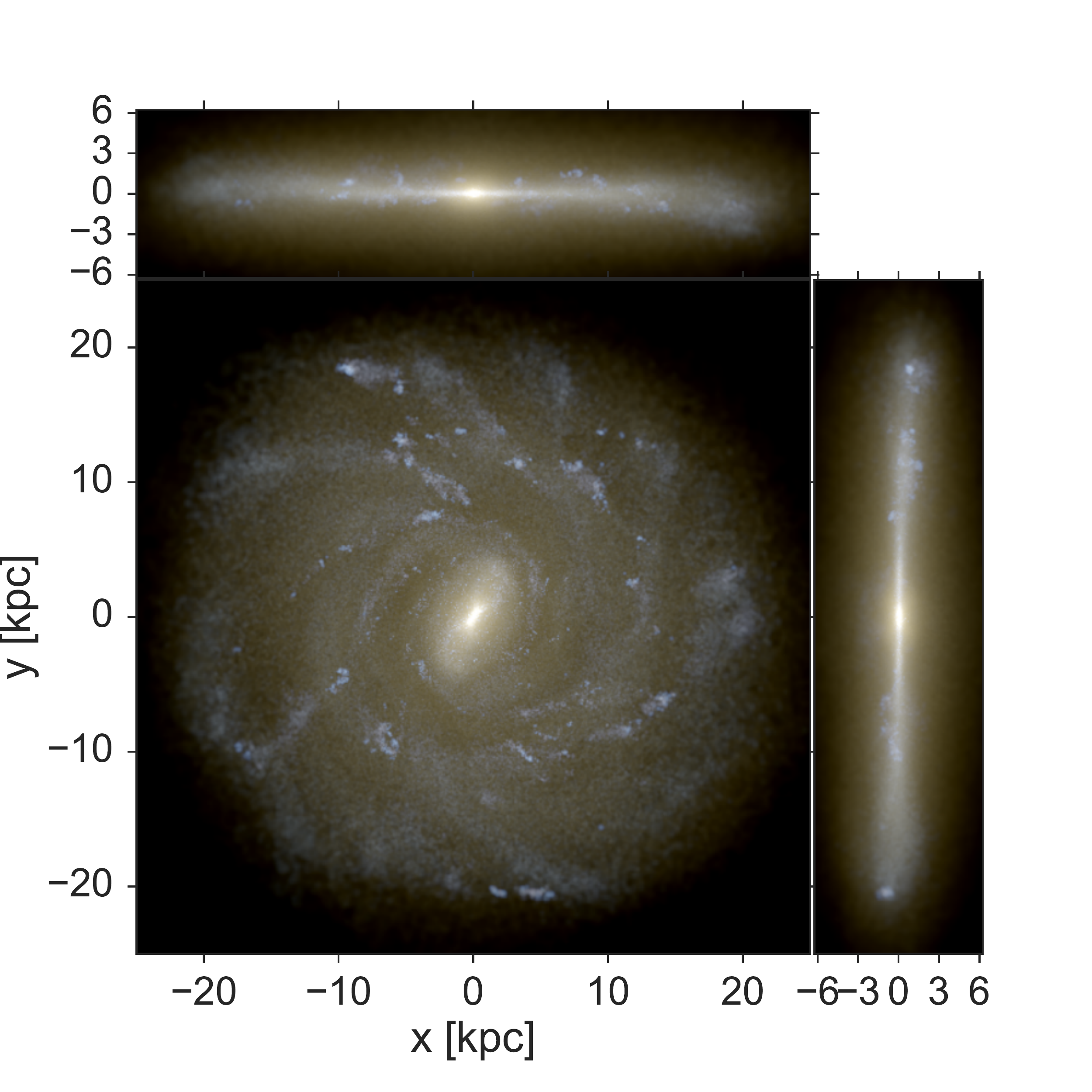}
\vspace{-.35cm}
\caption{
Stellar composite image of the galaxy in face-on and edge-on projections.  We use $i,v,u$-band fluxes to create the r,g,b maps. The colors are based on luminosities found using Padova Simple Stellar populations from Girardi and Marigo \citep{Girardi2010,Marigo2008}.  We did not run a radiative transfer code to account for dust attenuation.}
\label{fig:rgb}
\end{figure}

\section{Simulation} \label{sec:simulation}

The simulation analyzed in this work is a higher-resolution version of the galaxy g2.79e12 from the Numerical Investigation of a Hundred Astronomical Objects (NIHAO) project \citep{Wang2015}. The hydrodynamics, star formation recipes and feedback schemes exploited are the same as for the original NIHAO runs, which we summarize below.

Galaxies from the NIHAO sample have been proven to match remarkably well many of the properties of observed galaxies. This includes results from abundance matching \citep{Wang2015}, the local velocity function \citep{Maccio2016}, metal distribution in the Circum Galactic Medium \citep{Gutcke2016} and the properties of stellar and gaseous discs \citep{Obreja2016, Dutton2016b} or the morphological properties of high mass galaxies at high redshift ($z\sim 0.5 - 3$) \citep{Buck2017}. 
Therefore this galaxy is well studied and because it has a strong bar it is well suited to investigate the kinematical and morphological properties of a bulge system similar to the MW's bulge. An impression of the galaxy's face-on and edge-on projections is given in Fig. \ref{fig:rgb}.

\subsection{Hydrodynamics}

This high-resolution simulation was run with a modified version of the smoothed particle hydrodynamics (SPH) code {\texttt{GASOLINE2.0}} \citep{Wadsley2017} which includes substantial updates to the hydrodynamics as described in \citep{Keller2014} to alleviate known shortcomings of the SPH method \citep{Agertz2007}. The  modifications of the hydrodynamics improve multi-phase mixing and remove spurious numerical surface tension \cite{Ritchie2001}. We adopted a metal diffusion algorithm between particles as described  in \cite{Wadsley2008} and the treatment of artificial viscosity has  been modified to use the signal velocity as described  in \cite{Price2008}. Furthermore, the \cite{Saitoh2009} timestep limiter was implemented and \texttt{ESF-GASOLINE2} uses now the  Wendland C2 smoothing kernel  \citep{Dehnen2012}  to   avoid  pairing   instabilities.

Gas cooling is implemented via hydrogen, helium, and various  metal-lines as described  in \cite{Shen2010} and cooling functions are calculated using \texttt{cloudy}  \citep[version 07.02;][]{Ferland1998}. Furthermore the effects of  photo  heating and ionization from  the \cite{Haardt2005} UV background  and from Compton cooling in a temperature range from $10$ to $10^9$ K are included.

\subsection{Star Formation and Feedback}

The star formation recipe in the simulation follows the one described in \cite{Stinson2006}. Dense and cool gas ($n_{\rm  th}  >  10.3$ cm$^{-3}$, T $< 15,000$K) is eligible to form stars reproducing the Kennicutt-Schmidt Law.  The threshold number  density $n_{\rm th}$ of gas is set to the maximum density at which gravitational instabilities can be resolved in  the  simulation:  n$_{\rm th}$  $=$  50$m_{\rm  gas}/\epsilon_{\rm
  gas}^3 = 10.3$ cm$^{-3}$, where $m_{\rm gas}$ denotes the gas particle mass, $\epsilon_{\rm gas}$ the gravitational  softening of the gas and the value of 50 denotes the number of neighboring particles.

Two modes of stellar feedback are implemented as described  in \cite{Stinson2013}. The  first  mode  models the energy  input from  stellar  winds and  photo ionization from  luminous young  stars and happens  before any  supernovae explode. The energy input for this mode consists  of the total  stellar flux,  $2 \times  10^{50}$  erg of  thermal energy  per $M_{\odot}$ of the entire stellar population and the efficiency parameter for the  coupling of the  energy input  is set to  $\epsilon_{\rm ESF}
=13\%$ \citep{Wang2015}.

The second mode models the energy input from supernovae and starts 4 Myr after the formation of the star particle. It is implemented using the blastwave formalism as described in \cite{Stinson2006} and applies a delayed cooling formalism for particles inside  the blast region to avoid the artificial fast energy loss of the feedback energy in the dense regions of the interstellar gas surrounding the supernovae explosions due to its efficient cooling. See \cite{Stinson2013} for further information and an extended feedback parameter search. 

\subsection{Galaxy properties}
\label{sec:G_props}

\begin{table*}
\begin{center}
\caption{Properties of the g2.79e12 galaxy: \normalfont{total virial mass, $M_{200}$, virial radius, $R_{200}$, dark matter mass, $M_{\rm dark}$, total stellar mass within 0.1$R_{200}$, $M_{\rm star}$, total gas mass, $M_{\rm gas}$, mass of gas within 0.1R$_{200}$, $M_{\rm gas}^{\rm gal}$, disk scale length, $R_{\rm d}$, disc scale height, $H_{\rm z}$, circular velocity at 8 kpc, $V_{\rm circ}$ and radial extension of the bar, $L_{\rm bar}$.}}
\begin{tabular}{c c c c c c c c c c}
		\hline\hline
		$M_{200}$ & $R_{200}$ & $M_{\rm dark}$ & $M_{\rm star}$ & $M_{\rm gas}$ & $M_{\rm gas}^{\rm gal}$ & $R_{\rm d}$ & $H_{\rm z}$ & $V_{\rm circ}$ & $L_{\rm bar}$\\
		{[}$10^{12}$\Msun] & [kpc] &  [$10^{12}$\Msun] & [$10^{10}$\Msun] & [$10^{10}$\Msun] & [$10^{10}$\Msun] & [kpc]  & [kpc] & [km/s] & [kpc] \\
		\hline
		3.13 & 306 & 2.78 & 15.9 & 18.5 & 4.93 & 5.0 & 0.5 & 326.0 & 4.0\\
        \hline
\end{tabular}
\end{center}
\label{tab:props}
\end{table*}

\begin{figure}
\includegraphics[width=\columnwidth]{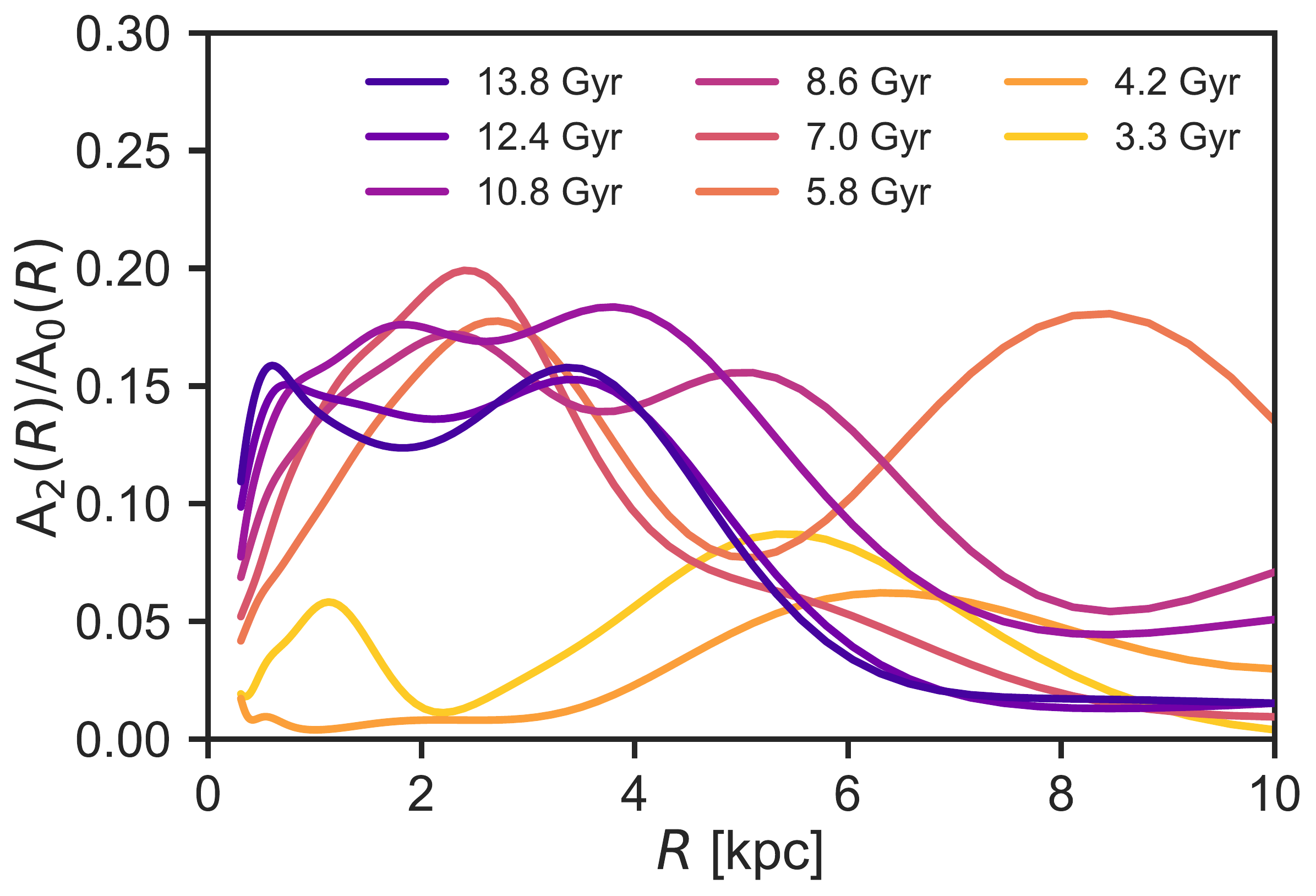}
\vspace{-.35cm}
\caption{Bar strength as a function of radius as calculated from the ratio of the fourier components $A_2/A_0$ for different times in the simulation. A prominent bar forms around $\sim7$ Gyr after the big bang at a redshift $z\sim1$.
}
\label{fig:bar_radius}
\end{figure}

This galaxy has been run using cosmological parameters from the \cite{Planck}, namely: \OmegaM=0.3175,   \OmegaL=0.6825,   \Omegab=0.049,  H${_0}$   =   67.1\kms\Mpc$^{-1}$, \sig8  = 0.8344. The mass resolution of this simulation is $m_{\rm dark}=5.1\times10^5 \Msun$ for dark matter particles and $m_{\rm gas}=9.4\times10^4 \Msun$ for the  gas particles. The initial star particle mass is set to $1/3\times m_{\rm gas}=3.1\times 10^4\Msun$. The corresponding force softenings are $\epsilon_{\rm dark}=620$ pc for the dark matter particles and $\epsilon_{\rm gas}=\epsilon_{\rm star}=265$ pc for the gas and star particles (see also table \ref{tab:sims}). However, the smoothing length of the gas particles (the scale of hydrodynamical forces)  can be much smaller, e.g.  as  low  as  $h_{\rm  smooth}\sim 20$ pc.

The main properties of this galaxy can be found in table 2. The final total mass within the virial radius (R$_{\rm vir}\sim 300$ kpc) is  M$_{\rm tot}=3.13\times 10^{12}$\Msun and the stellar mass of the galaxy (measured within $0.1\times$ R$_{\rm vir}$) is M$_{\rm star}=1.59\times10^{11}$\Msun. The galaxy's stellar disk has a scale length of R$_{\rm d}\sim 5$ kpc and a total scale height of H$_{z}\sim500$ pc within the innermost 5 kpc and  H$_{z}\sim1$ kpc in the outskirts at  $R>10$ kpc.

Before we turn our attention to the results of a detailed analysis of our simulation let us spend a few words on the bar of our simulation. Similar to the MW, g2.79e12hr contains a bar. It is important to have a good description of this bar because there is some confusion in the literature as to what we call the bulge of the MW. Some authors use the terminology pseudobulge some use boxy/peanut bulge but all of them mean the same thing. The structure we see in the center of our MW is the bar, and this is the same as the bulge (when viewed edge on,  which appears to be boxy/peanut shaped. 
Therefore, throughout the rest of the paper we will use (boxy/peanut) bulge or bar for referring to the same thing. If we take all stars in the inner region, including the disk surrounding the bar, we refer to the inner most region of the MW. At redshift zero the bar in the simulation extends out to about 4 kpc from the galaxy center. We characterize the bar in our simulation using the $m=2$ Fourier modes of the galaxy:
\begin{equation}
A_2 = \sum_j \exp\left(i2\varphi_j\right) m_j
\end{equation}
where $m_j$ and $\varphi_j$ are mass and azimuth angle of the stars. The sum extends over all stars in the considered region (spherical bins) of the galaxy. This $m=2$ mode encodes the bar strength defined as
\begin{equation}
A_2/A_0 = \frac{\left|A_2\right|}{\sum_jm_j}
\end{equation}
We use this quantity to estimate the formation time of the bar as a function of radius (see Fig. \ref{fig:bar_radius}). In Fig. \ref{fig:bar_radius} we show the bar strength $A_2/A_0$ as a function of radius at several redshifts represented by the different colors. After an initial strong fluctuation of the bar strength due to a violent merger dominated phase before redshift $z=2$, the bar strength grows continuously from redshift $z=2$ onwards. Before $z=1$ the value of $A_2/A_0$ in the innermost 4~kpc is relatively low, while after that time it shows a larger value,  $A_2/A_0\sim$0.17. This is the signature of a bar extending up to $\sim4$ kpc. The bar in this simulation is first clearly visible in surface density images at redshift $z=1.3$ or $\sim8$ Gyr ago,  and from then onwards it grows in strength and size. At redshift $z\sim0.75$ or equivalently $\sim$6.5 Gyr ago, the bar buckles forming the boxy/peanut bulge. This process causes a reduction in bar strength and an increase in vertical velocity dispersion. Thus we conclude that  the bar in our simulation formed between $\sim$10 and 8 Gyr ago at a redshift of $z\sim2-1$ as can be confirmed by visual inspection.

\section{Bulge properties: Comparison to the Milky Way}
\label{sec:B_props}

\begin{figure}
\includegraphics[width=\columnwidth]{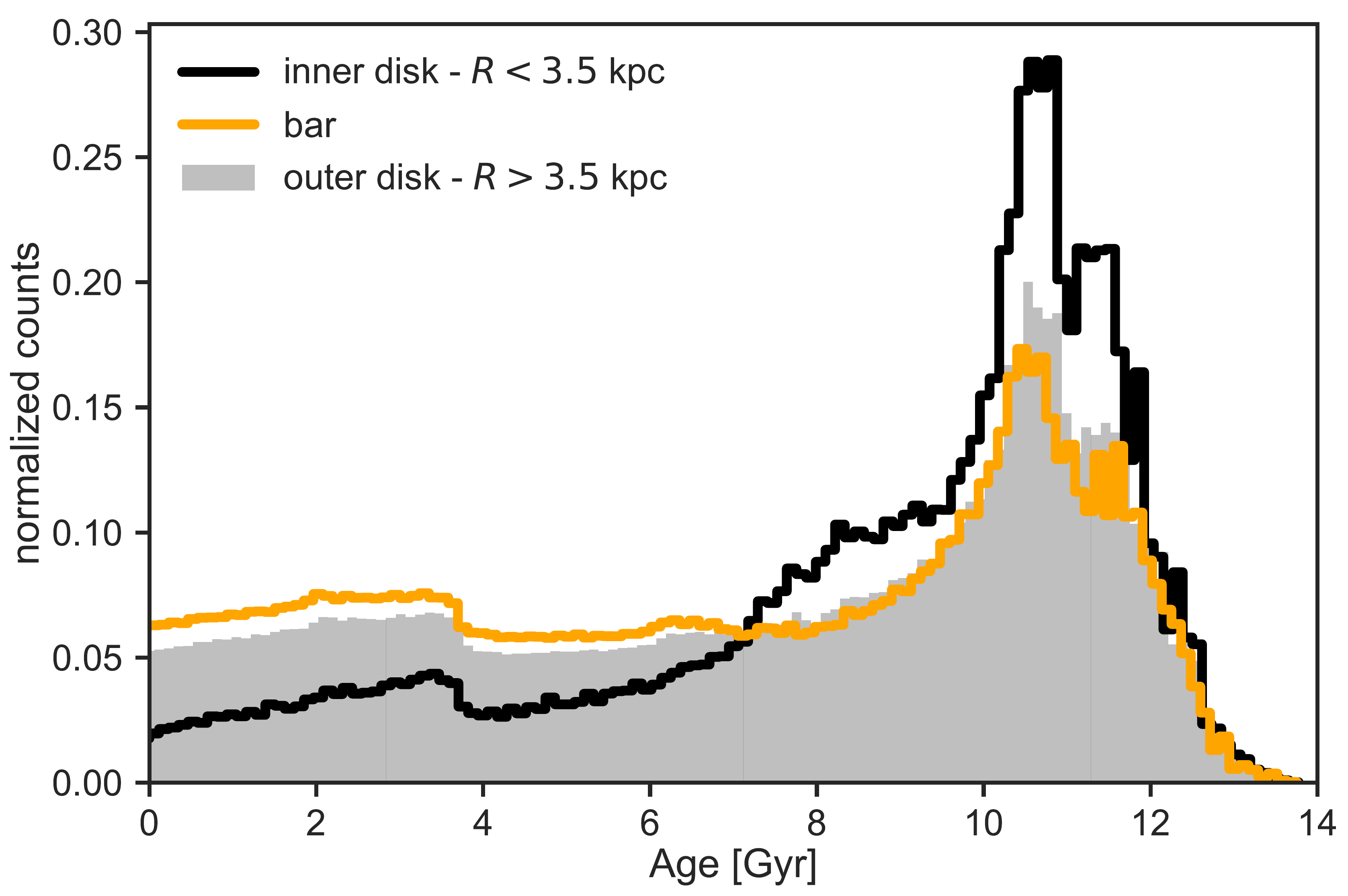}
\vspace{-.35cm}
\caption{Age distribution of stars in the bar (orange histogram), the inner disk ($R<3.5$ kpc, black histogram) and the outer disk ($R>3.5$ kpc, gray filled histogram) in the galaxy at redshift $z=0$. Stars belonging to the bar meet the following spatial selection criteria: $-3.5<x/\rm{kpc}<3.5$, $-1.25<y/\rm{kpc}<1.25$ and $-1.0<z/\rm{kpc}<1.0$ while the inner disk is defined as the inner most $3.5$ kpc of this galaxy excluding the bar. 
}
\label{fig:age_dist_all}
\end{figure}

\begin{figure*}
\includegraphics[width=.33\textwidth]{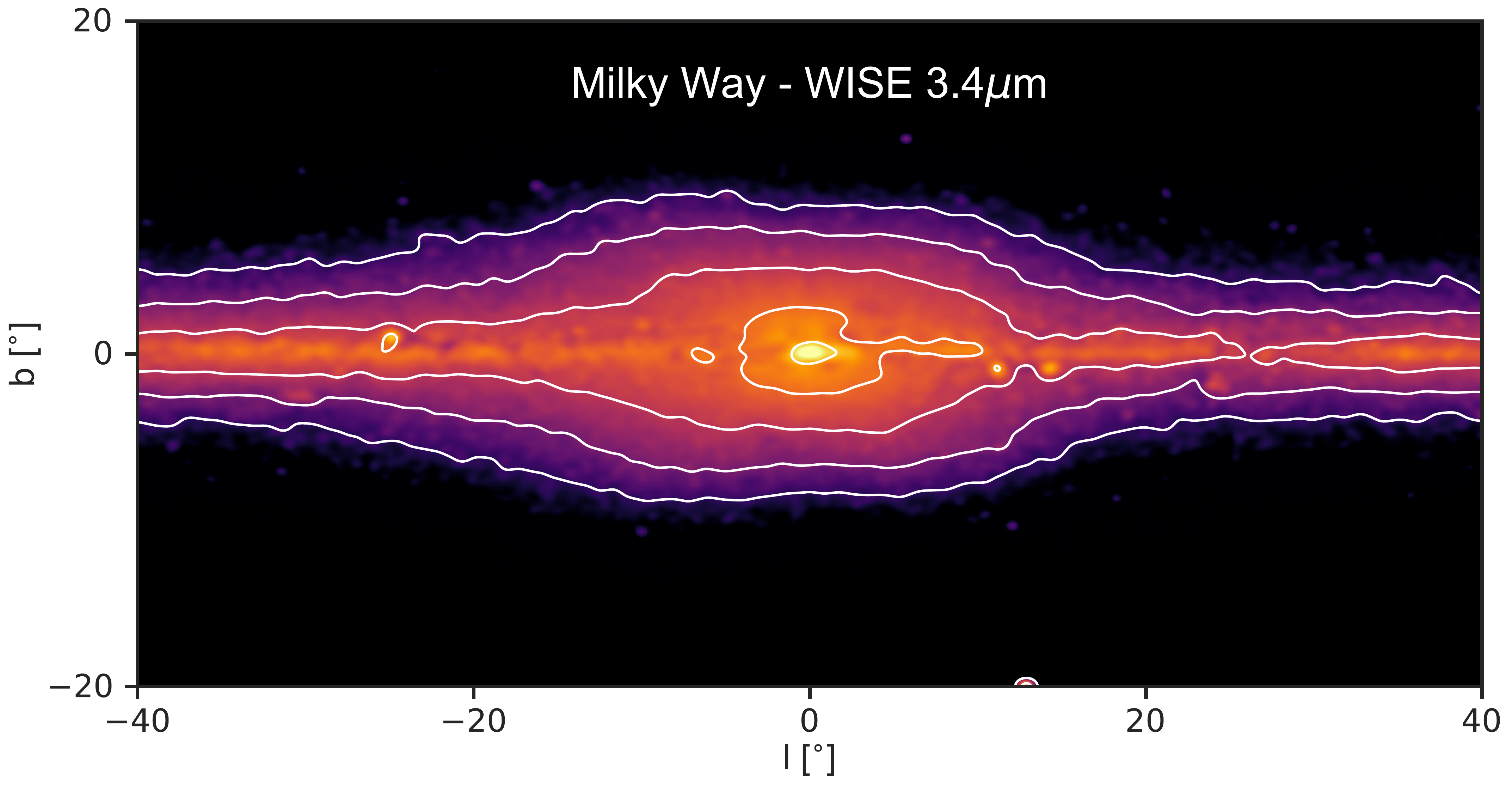} 
\includegraphics[width=.33\textwidth]{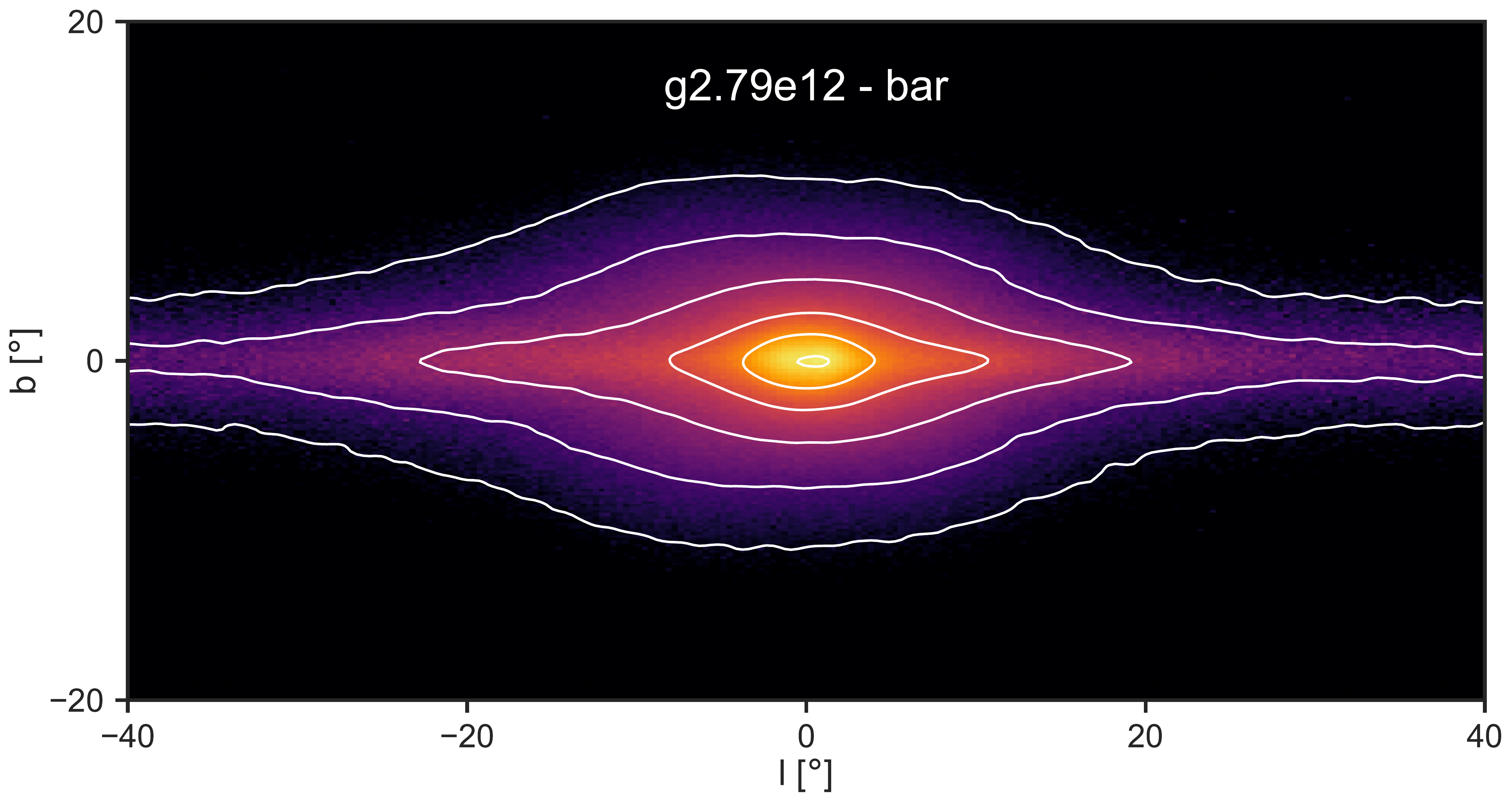}
\includegraphics[width=.33\textwidth]{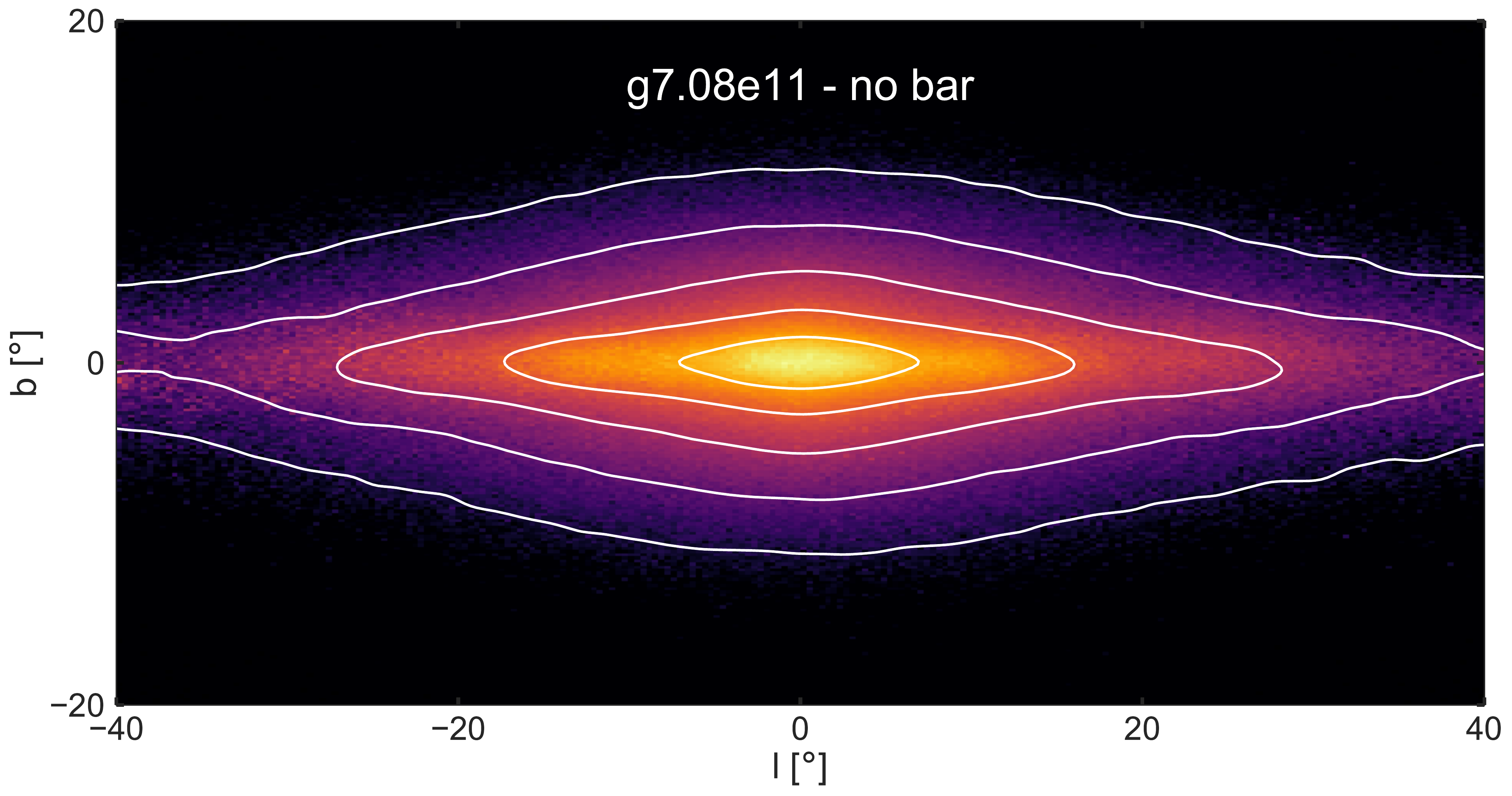}
\vspace{-.35cm}
\caption{\emph{Left panel:} Milky Way $3.4\mu$m image of the WISE satellite \citep[][]{Ness2016b}. \emph{Middle panel:} K-band image of the boxy/peanut bulge of the g2.79e12 simulation in galactic coordinates. \emph{Right panel:} K-band image of the bulge of the non-barred simulation g7.08e11 (Buck et al. 2018 a subm.).  The two left most panels show clearly the boxy/peanut shape of the bulge region with a qualitative visual similarity between the MW and the simulation analyzed here. In contrast, the right most panel shows one of our non-barred galaxies clearly showing morphological differences with respect to the other two panels.}
\label{fig:wise_comp}
\end{figure*}

In this section we compare the properties of the bulge region of our simulation to key observations from our own Galaxy to establish the similarity between the simulation and the MW. In the analysis that follows we place the Sun at (x,y,z)=(8,0,0) kpc and rotate the simulation such that the bar is inclined at $27^\circ$ with respect to the line-of-sight to match the position of the Sun in the MW \citep{Wegg2013}. We then transform all coordinates to galactic longitude and latitude ($l,b$).  

In observations the metallicity of individual stars has been used to infer the temporal evolution of the bulge structure (i.e. metal poor stars are mostly old, metal rich stars are mostly young). Ages for small samples of individual stars have been measured using microlensing events \citep[see][]{Bensby2017}. 
For several reasons, we use in this work the stellar ages of our star particles directly and not the metallicity of the stellar particles from our simulation.  First, we find for this particular simulation that stellar populations show a large scatter in ages for single metallicity bins (see e.g. Fig.  \ref{fig:SFR} in the Appendix), similar to the results of \cite{Minchev2017}. This prevents a clean investigation of the temporal build up of structures in our simulations using stellar metallicity. Second, the fundamental variable to measure galactic formation and evolution is age, and it is straightforward to obtain in simulations. Observationally, stellar ages are now being measured from spectra of bright red giant stars in the bulge \citep{Ness2016, Martig2016}. Therefore, our choice for stellar ages allows for a direct link to new generations of spectroscopic surveys [e.g. MOONS \citep{C2012}, APOGEE-2 \citep{Zasowski2017} and Sloan V \citep{Kollmeier2017}].

\subsection{The age distribution of bulge stars}

Fig. \ref{fig:age_dist_all} shows the age distribution of the stars in our model galaxy, for stars of the disk within $25$ kpc from the galactic centre and a height smaller than $6$ kpc from the mid plane. Three spatial bins are shown, (i) stars in the inner galaxy,  $<3.5$ kpc of the galactic center but excluding the bar, (ii) stars in the bar selected to meet the spatial selection criteria $-3.5<x/\rm{kpc}<3.5$, $-1.25<y/\rm{kpc}<1.25$ and $-1.0<z/\rm{kpc}<1.0$, (for the bar aligned with the x-axis) and (iii) stars of the disk outside of the inner $3.5$ kpc. Our model galaxy shows a wide distribution of stellar ages ranging from $0$ to $\sim14$ Gyr with a peak around $10-11$ Gyr and a tail towards lower ages with a slight peak for very young stars ($<3$ Gyr). The distribution of stellar ages in the outer disk and in the bar is very similar. However, the inner disk excluding the bar shows a lower (larger) proportion of young (old) stars. This points towards a preferential origin of bar stars from the outer disk which we study in much more detail in a follow-up paper (Buck et al. 2018 subm.). We will come back to the similarity of ages in the inner disk region and those trapped in the bulge in Section \ref{sec:origin}, when we discuss the origin of stars in the bar.

Comparing our results to observed age distributions for stars in the MW and its bulge we find very good agreement. The total age distribution of stars in this model galaxy is very similar the reconstructed one of the MW by \citet{Snaith2015} using spectroscopic data from \citet{Adibekyan2012}. Their age distribution shows a strong peak at $\sim11$ Gyr and a tail towards lower ages with indications of a secondary peak at $\sim3$ Gyr well in agreement with our simulation.

Using the APOGEE data set \citep{Apogee}, \citet{Zhou2017} find a wide distribution of ages in the bulge region of the MW. Their high metallicity population shows ages ranging from $2$ to $14$ Gyr while their metal poor population is slightly older with ages between $6$ and $14$ Gyr. This is consistent with the  findings of \citet{Bensby2013} and \citet{Bensby2017}. The age-metallicity relation reconstructed from the APOGEE data is quite flat for stars in the bulge region, indicating a wide range of stellar age populations at similar metallicities. This suggests the existence of multiple stellar populations in the bugle region, ranging from young to old. This is in very good agreement with the results obtained from our simulation. 

On the contrary, using ages derived from proper motion cleaned color magnitude diagrams the fraction of young stars in the bulge region of the MW is found to be less than 3.5\% \citep{Clarkson2008,Clarkson2011,Gennaro2015}. This discrepancy with the results of e.g. \citet{Bensby2013} lead \citet{Haywood2016} to suggest that an age-metallicity degeneracies might make a young population undetectable using color magnitude diagrams \citep[see recent review by][for a more detailed discussion]{Barbuy2018}.

\subsection{Morphology: The visible X-shape and the split in stellar counts}

\begin{figure}
\includegraphics[width=\columnwidth]{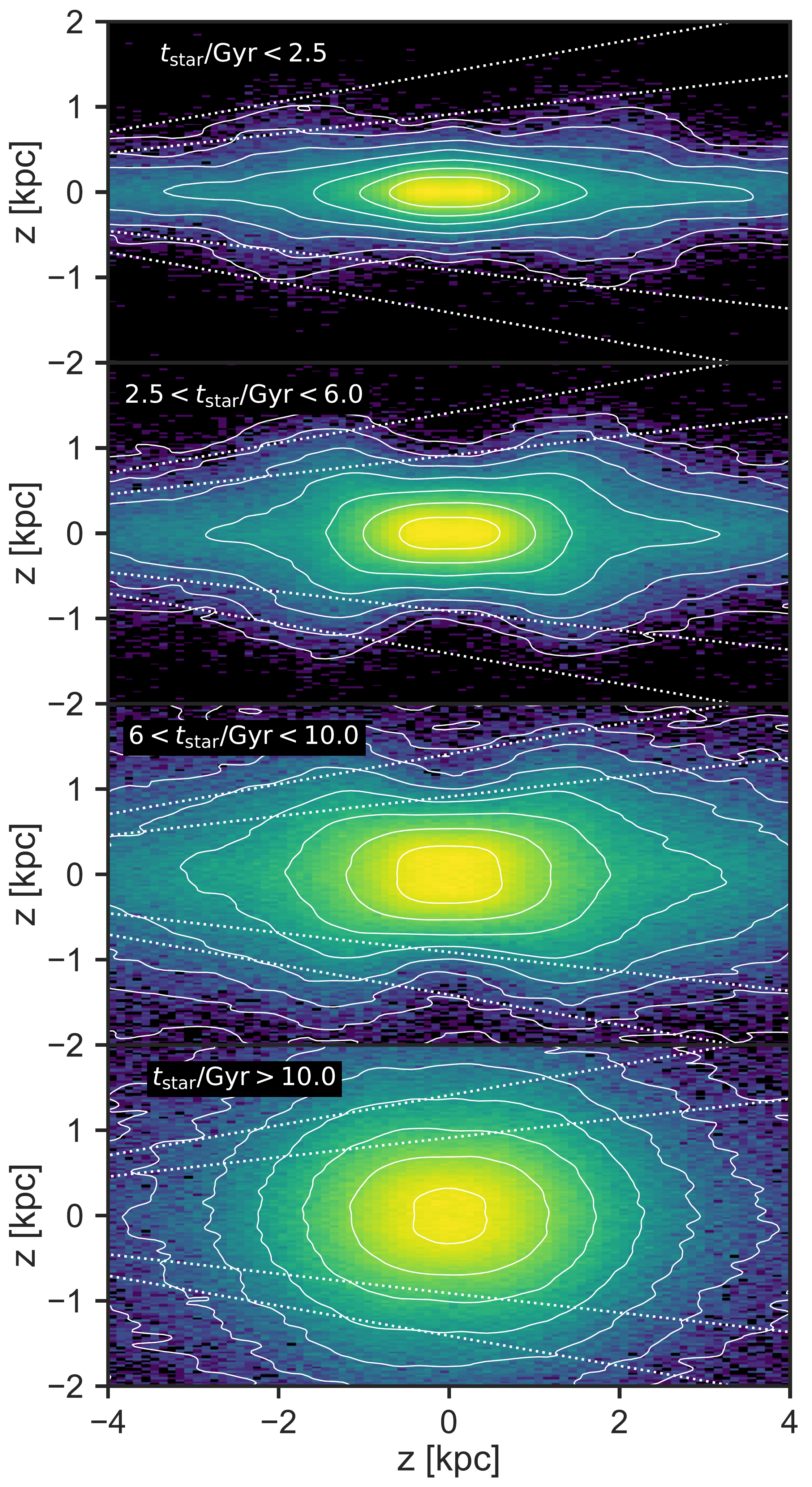}
\vspace{-.35cm}
\caption{Side on surface density plots of  a thin slice of thickness  $1$ kpc centered on $y=0$ kpc of the X-shaped bulge in our simulation. The sun's position is assumed to be at $(x,y,z)=(8,0,0)$ kpc. The white dashed lines indicate the cones in which we do star counts along the line of sight for Fig. \ref{fig:split}. From top to bottom we show  the different age populations used  in Fig. \ref{fig:split} from the youngest stars (age$<2.5$ Gyr)  to the oldest stars (age$>10.0$ Gyr).
}
\label{fig:x-shape}
\end{figure}

\begin{figure}
\includegraphics[width=\columnwidth]{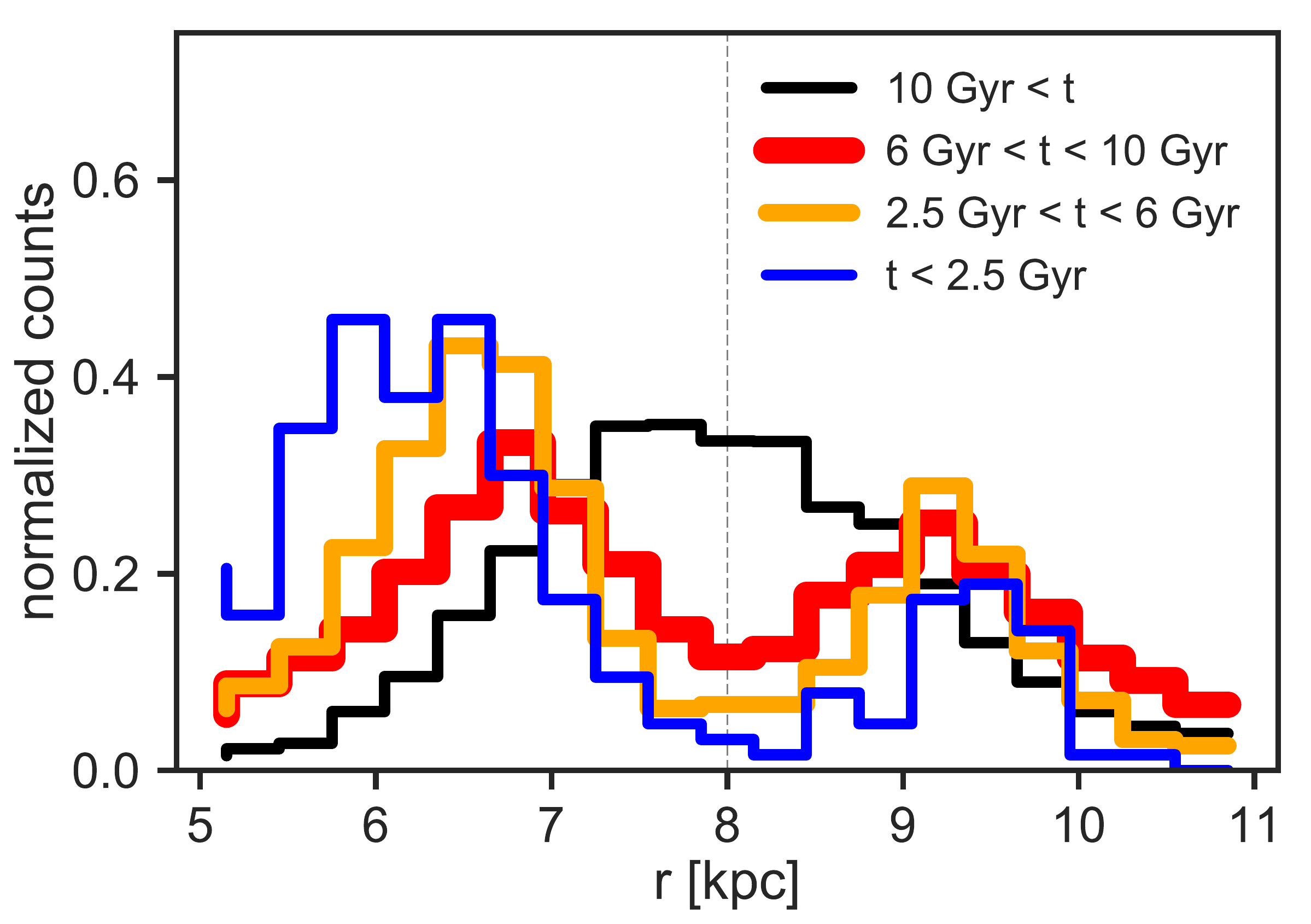}
\vspace{-.35cm}
\caption{
Star counts as a function of distance from the sun in different age bins for line-of-sights going through the center of the galaxy ($-2.0^\circ<l<2.0^\circ$) at a height of $6.5^\circ<\vert b\vert<10^\circ$ above the galactic plane similar to ARGOS observations of our Milky Way \citep{Ness2013,Ness2013a}. The Galactic center position is indicated by the vertical dashed gray line at $r=8$ kpc.}
\label{fig:split}
\end{figure}

\begin{figure*}
\includegraphics[width=0.5\textwidth]{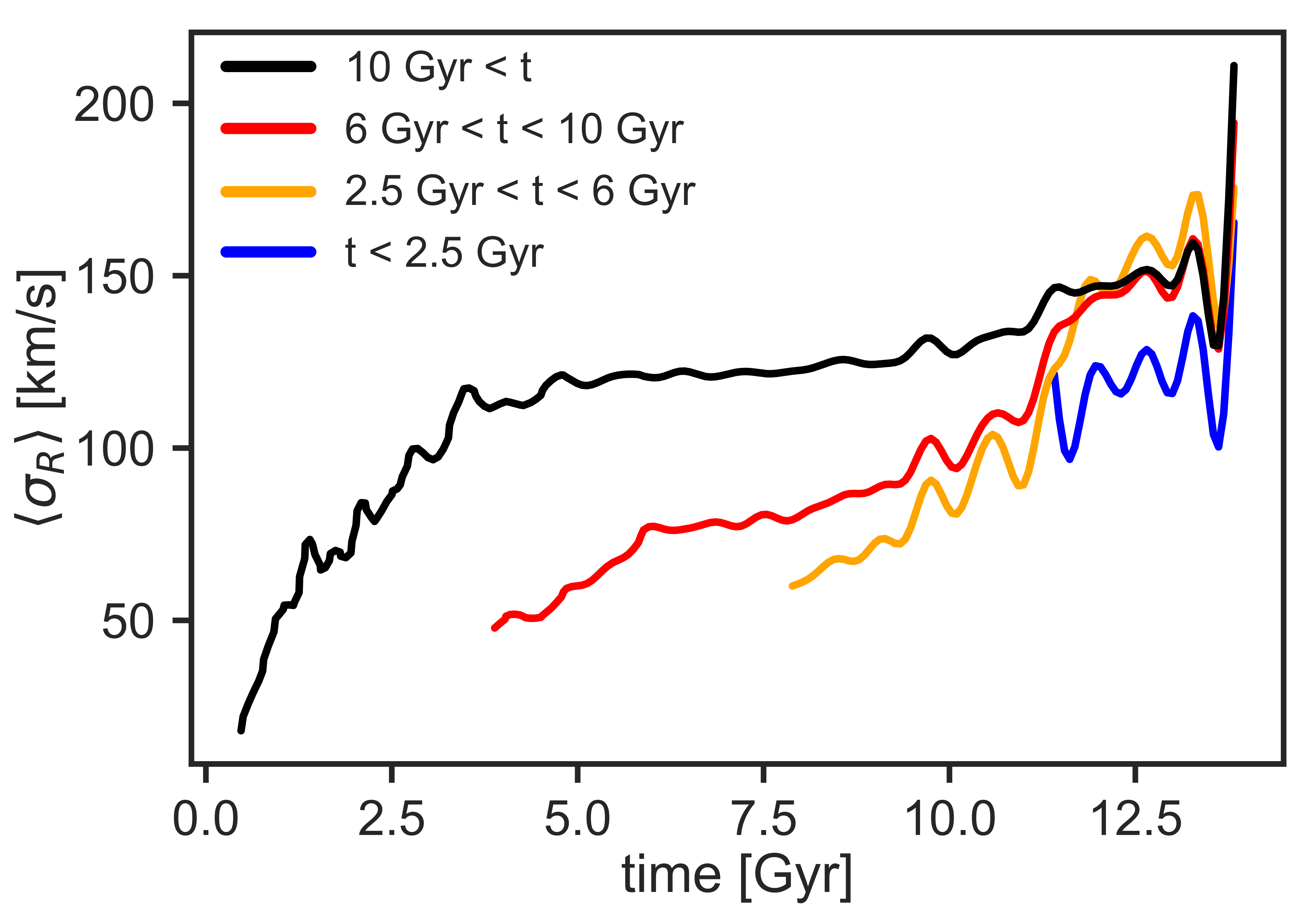}
\includegraphics[width=0.5\textwidth]{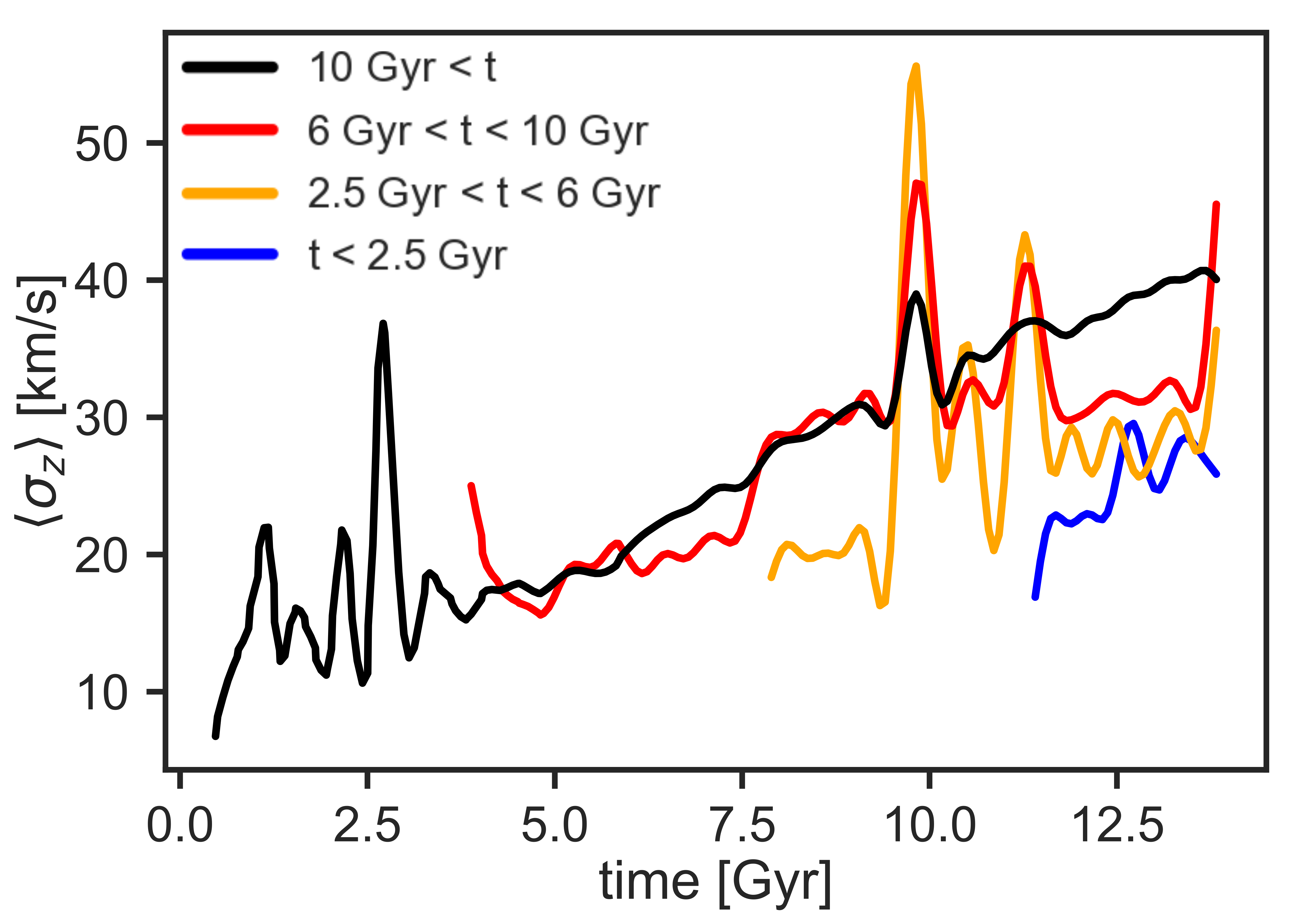}
\vspace{-.35cm}
\caption{Evolution of the radial (left panel) and vertical (right panel) velocity dispersion of stars in the inner region ($R<3.5$ kpc).  Color coding is the same as in Fig. \ref{fig:split} with  blue lines showing stars younger than 2.5 Gyr,  orange lines showing stars with $2.5 <$ age/Gyr $< 6$, red lines $6 <$ age/Gyr $< 10$ and black showing the oldest stars with age $>10$ Gyr. }
\label{fig:fractionation}
\end{figure*}

The MW contains a  boxy/peanut shaped bulge with a strong X-shaped structure clearly visible in photometric imaging. In \citet{Ness2016b} the X-shape of the bulge is readily seen from the WISE satellite photometry, which penetrates the dust obscured inner-most region. This X-shape and the overall boxy/peanut bulge morphology are not uncommon in extragalactic spirals \citep{Bureau2006,Gonzalez2017}.  

In Fig. \ref{fig:wise_comp} we show  a K-band image of our simulation side-by-side to the WISE $3.4  \mu$m image with data taken from \cite{Ness2016b} to show the remarkable qualitative similarity  between the morphology in the simulation and in our own Galaxy.
A feature of the presence of a peanut-shaped bulge is the X-shaped structure which can often be seen if viewed side-on. In Fig. \ref{fig:x-shape} we show this feature for the different stellar populations of the simulation. This figure shows the surface density of stars in a thin slice of 1 kpc thickness centered around the peanut bulge mid-plane. Starting from the youngest stars the strength of the X-shape gets stronger with age. Only the oldest stars do not show this feature. The X-shaped morphology gets less prominent and more boxy, or thicker (stars reach larger heights above the galactic mid-plane) for increasingly older populations.  The X-shaped morphology is strongest for the intermediate age population ($2.5 < t_{\rm star}/\rm{Gyr} < 6.0$), where the stars extend in the arms of the X to the highest latitudes and trace orbits down to the lowest latitudes at the very center. The respective mass fractions in the innermost 4 kpc in the four different age bins in terms of total stellar mass of this simulated galaxy are: 10\%, 12\%, 18\% and 18\% for stellar particles in the age bins $t_{\rm star}/\rm{Gyr} < 2.5$, $2.5 < t_{\rm star}/\rm{Gyr} < 6.0$, $6.0 < t_{\rm star}/\rm{Gyr} < 10.0$ and $t_{\rm star}/\rm{Gyr} > 10.0$.

The bulge of our own Galaxy can only be observed from within the Galaxy. Thus to reconstruct the structure of the bulge one has to rely on line-of-sight counts of stars. The X-shaped structure of the bulge thus translates into a double peaked distribution of stars as a function of distance. This feature was first observed in our own Galaxy from photometry in the star counts along the line of sight \citep{McWilliam2010, Nataf2010} and found from spectroscopy to be metallicity dependent e.g. \citep{Ness2012, Uttenthaler2012, RJ2014}. For the MW there have been published various studies of star counts for sight-lines going through the center of the Galaxy \citep[$-2.0^\circ<l<2.0^\circ$ at  $6.5^\circ<\vert b\vert<10^\circ$ above the galactic plane, e.g.][]{Ness2013,Ness2013a}. We use the same sight-lines in our simulation which are indicated by the dashed white lines in Fig. \ref{fig:x-shape}. From this we see that the anisotropic distribution of stars in the X-shaped structure lead to a double peaked distribution of star counts. In Fig. \ref{fig:split} we show the radial distribution of star counts in the bulge region of our simulation.  Similarly to the observations of the MW bulge we count the number of stars as a function of distance in the above mentioned sight lines. We restrict ourselves to distances ranging from 5-11 kpc from the Sun's assumed position (thus $\pm$ 3 kpc from the galactic center) and divide our stars into four different age bins.

The resulting split in the stellar line-of-sight counts  of stars in the simulation looks qualitatively very similar to the observations of the ARGOS survey \citep{Ness2013,Ness2013a}. Up to a stellar age of 10 Gyr we see a split in the radial distribution of stars and we checked that for any age bins older than 10 Gyr we do not see a split. Therefore we conclude that this population older than 10 Gyr is not part of the boxy/peanut structure. In Fig. \ref{fig:split} it is clearly visible that the peak separation becomes smaller for progressively older stellar populations. This is similar to what is observed for the MW -- as a function of stellar metallicity -- where the metal poor population is less separated compared to the metal rich population. This is confirmed by the surface-density maps shown in Fig. \ref{fig:x-shape}.  The slight asymmetry of the line-of-sight counts for the younger stellar populations visible in  Fig. \ref{fig:split} can be explained by the inclination of the cones use to count the stars (see white dashed lines in Fig. \ref{fig:x-shape}). On the near side the cones are close enough to the mid-plane to cut right through the X-shape while on the far side the cones are to far from the mid-plane and the X-shape does not extend high enough above the mid-plane to fully intersect with the cones. 

The question of the origin of such a spatial separation of different populations of stars was recently addressed by \cite{Debattista2016} using pure $N$-body simulations and an isolated simulation of galaxy formation.  These authors find that initially co-spatial stellar populations with different in-plane random motions  separate  when a bar forms. In their simulations the stellar population with higher radial velocity dispersion becomes a vertically thicker box while the radially cooler population stays thinner and forms a peanut-shaped  bulge. These authors termed this mechanism \textit{kinematic fractionation}. One important prediction of this mechanism is that even after the buckling instability of the bar, the disc stars can be scattered by the bar to large heights above the disc. As explained in \citet[][section 2]{Debattista2016}: \textit{as long as the bar is slowing down, therefore, the disc will continue to thicken at different rates for different radial dispersion populations, allowing the separation of populations to persist in subsequent evolution.} This mechanism is important in this simulation as we see the bar forming around 8 Gyr ago. We also observe a strong X-shape in the stars with ages between 2.5 and 6 Gyr as can be seen in Fig. \ref{fig:x-shape}. This however points towards further dynamical influences after bar formation acting on the redistribution and star formation in the already barred galaxy. E.g. \citet{Fragkoudi2017} have shown that thin disc stars get trapped more easily in the bar compared to thick disc stars which explains the enhanced contribution of young stars to the bar compared to the inner disc seen in Fig. \ref{fig:age_dist_all}.

We confirm with this fully cosmological simulation that indeed the radial velocity dispersion is higher for older stellar populations over the whole cosmic time of evolution which agrees with the picture presented in \cite{Debattista2016}. It is reassuring that the same mechanism is at work in pure $N$-body simulations and idealized, isolated simulations of galaxy formation as well as in the fully cosmological context. These mechanisms thus also seem to shape the galaxy in an environment where there are many additional perturbations, from incoming satellites and minor mergers,  as per this work. 

In Fig. \ref{fig:fractionation} we show the evolution of the radial (left panel) and vertical velocity dispersion (right panel) of the four different stellar age populations as a function of time. The color-coding of the lines is the same as for Fig. \ref{fig:split}.  For both panels we trace back all the stars within a sphere of $3.5$ kpc at redshift $z=0$ and we calculate the velocity dispersion in cylindrical bins aligned with the stellar disc and then average over all bins, as in \cite{Debattista2016}. We see that the lowest radial dispersion is found in the youngest stars (blue line), and that the radial velocity dispersion increases with increasing stellar age. The vertical velocity dispersion of the stars is much more similar among different sub-populations as can be seen from the right panel of Fig. \ref{fig:fractionation}. The increase in vertical velocity dispersion around 7.5 Gyr is due to the buckling instability of the bar, and the spike in vertical velocity dispersion at $t\sim10$ Gyr is due to a massive satellite passing through the stellar disc of our simulated galaxy causing almost equal extra heating of the stellar populations. Interestingly the radial velocity dispersion is less affected by the disc passage of the satellite.

As a last remark, we see gradual kinematical heating of all stellar populations in the simulation. With cosmic time the radial and vertical velocity dispersion of the stars increases, although the relative increase is stronger for the vertical velocity dispersion.

\subsection{Kinematics: Rotation and dispersion profiles as a function of stellar age}
\label{sec:bulge_kinematics}

\begin{figure*}
\includegraphics[width=\textwidth]{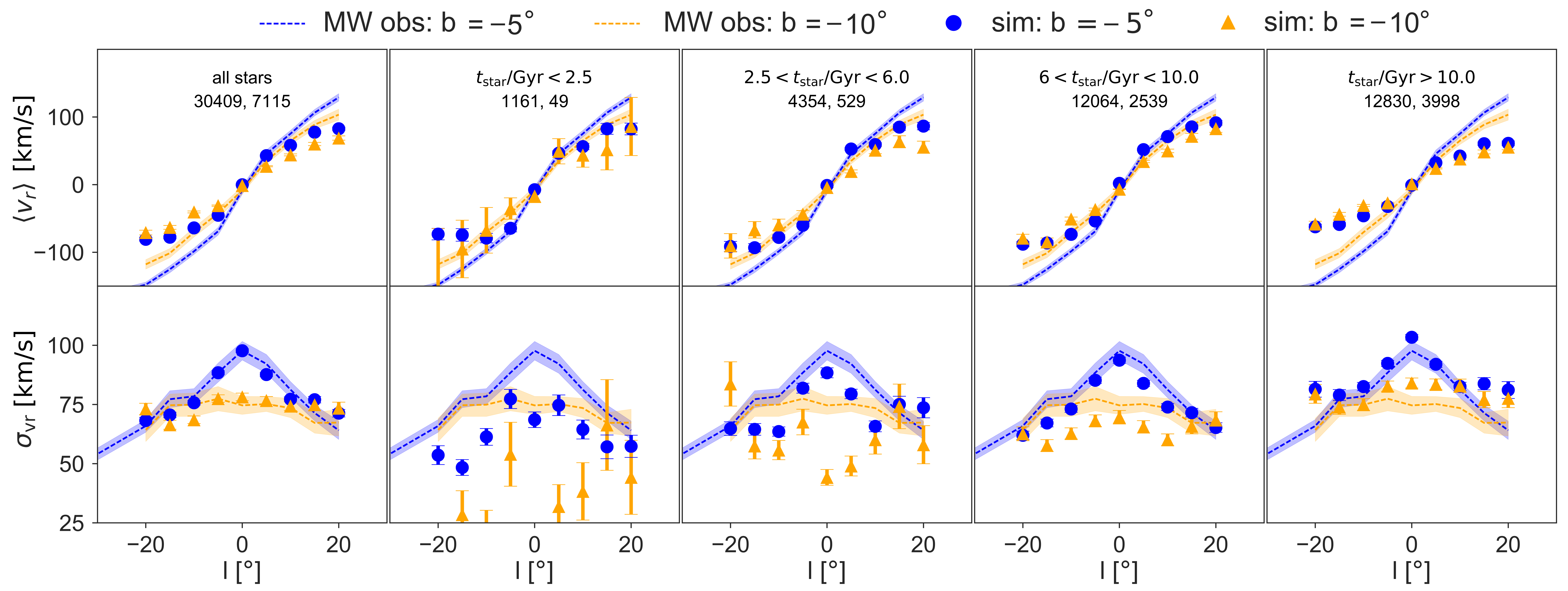}
\vspace{-.35cm}
\caption{
Rotation (top) and dispersion (bottom) profiles for stars of different ages for two different latitudes $\vert b\vert=5^\circ$ (blue) and $\vert b\vert=10^\circ$ (yellow).  Rotation is the mean radial velocity for stars at distances $5$ to $11$ kpc along the line-of-sight, as a function of galactic longitude $l$. The dispersion profile is the velocity dispersion along the line-of-sight as a function of galactic longitude $l$. Lines show the measurements from ARGOS \citep{Ness2013,Ness2013a} for the whole sample of MW stars and dots/triangles show the result obtained for our simulation. From left to right, the panels show the profiles for: all stars, young stars (age $< 2.5$ Gyr), stars with ages $2.5 <$ age/Gyr $< 6$, stars with ages $6 <$ age/Gyr $< 10$ and the oldest stars with age $>10$ Gyr. The numbers in each panel indicate the amount of stellar particles in each sample, the first for sight lines with $\vert b\vert=5^\circ$, the second for $\vert b\vert=10^\circ$.}
\label{fig:rot_disp}
\end{figure*}

\begin{figure*}
\centering 
\includegraphics[width=.85\textwidth]{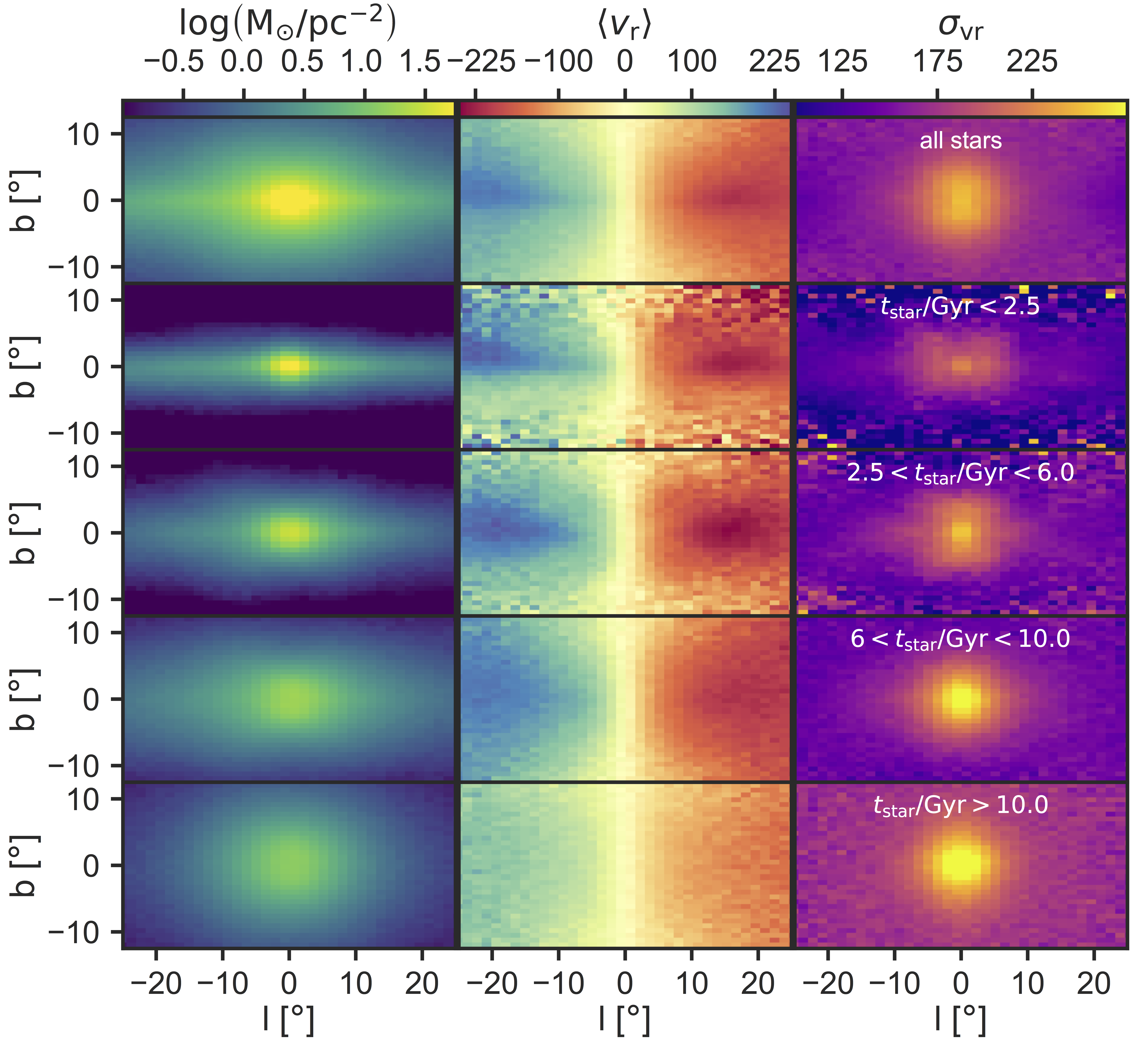}
\vspace{-.05cm}
\caption{Surface density (left column), mass-weighted rotation (middle column) and dispersion maps (right column) for different stellar age populations in $(l,b)$-projection for stars within $R=25$ kpc from the center of our simulated galaxy.  The \emph{first row} shows the results for all stars, the \emph{second row} shows stars younger than $2.5$ Gyr , the \emph{third row} shows stars in the age range $2.5$ Gyr up to $6$ Gyr, the \emph{fourth row} shows stars in the age range $6$ Gyr up to $10$ Gyr and the \emph{bottom row} shows stars older than $10$ Gyr.
}
\label{fig:maps}
\end{figure*}

The stars in the inner MW ($R<3.5$ kpc) show a distinct rotation and dispersion profile \citep{Kunder2012,Ness2013,Zasowski2016}. The rotation is the mean radial velocity for stars at distances $5$ to $11$ kpc along the line-of-sight, as a function of galactic longitude $l$ and the dispersion profile is the velocity dispersion along the line-of-sight as a function of galactic longitude $l$. In Fig. \ref{fig:rot_disp} we compare the rotation and dispersion profiles of the bulge stars in our simulation (for all stars and for our four different age populations) with observed profiles for the MW calculated from \texttt{ARGOS} data -- for all stars, across all [Fe/H]: this helps to guide the eye to compare how the trends in the simulation change with age. 

In the left panels of Fig. \ref{fig:rot_disp} we compare the rotation curves (upper panel) and the dispersion profiles (lower panel) for all stars in the observational sample and all stars in the simulation. Following the observations we calculate rotation and dispersion in  $2^\circ$ sized bins in $(l,b)$. Due to the higher mass of our model galaxy compared to estimated values of the MW, the rotation and dispersion values obtained from the simulation are slightly too high compare to the observed values. Thus, we rescale the rotation and dispersion values by a constant factor of $\sim0.45$ to match the observed dispersion value at $(l,b)=(0^\circ,-5^\circ)$ of the observations. This rescaling is valid since the interesting feature of the observations is not the absolute value of rotation or dispersion but the particular shape of the profiles. After the rescaling, the dispersion profile of all stars show excellent agreement with the observed profiles (colored shaded bands) for both latitude bins ($b=-5^\circ$ blue dots and $b=-10^\circ$ yellow triangles). Our simulation is able to recover the flat dispersion profile for large heights above the plane and the triangular peaked shape of the dispersion profile closer to the disc mid-plane. The rotation profile of our simulation shows the same qualitative behavior as the observations but somewhat smaller maximum values of rotation for the smallest and largest $l$-bins. The reason for this is most likely a slight miss match in size of our simulation and the MW.

The other eight panels of Fig. \ref{fig:rot_disp} show the rotation and dispersion profiles for stars  of our simulation in the different age bins (chosen to be the same as in Fig. \ref{fig:split}), together with the profiles of all stars in the observations to guide the eye (a closer, more direct comparison between simulation and MW will be done in a follow-up paper).  We do not see large differences in the rotation profiles of different age populations. Note however that the rotation is slowest for the oldest stars (the far right panel). This is in qualitative agreement with the observations given that metal-poor stars are generally younger than metal-rich stars. Observationally, the rotation profile shows only a very slight variation for different metallicity populations in our MW for stars [Fe/H] $>$ -1.0. \citep[see e.g.][]{Ness2013a}. However, for the most metal-poor stars observed in the bulge, that is the 5\% of stars with [Fe/H] $< -1.0$, which have the highest dispersions,  the rotation is far slower than the more metal rich stars (on the order of 50\% of the more metal rich population, as measured by ARGOS). The  RR Lyrae population, which peaks at [Fe/H] = -1.0 in the bulge, shows no rotation at all, in the observations of \citet{Kunder2016}. 

We now turn to examining the dispersion profiles which for the MW bulge stars show a strong variation for different metallicity populations \citep{Ness2013, Ness2016, Zasowski2016,  Babusiaux2016}. The simulation shows that with increasing age of the stars, the velocity dispersion increases for both $b$-bins. This is in qualitative agreement with what is seen in the ARGOS survey for stars of decreasing [Fe/H], down to [Fe/H] $>$ -0.5 \citep[see e.g.][Fig. 6]{Ness2013a}. Observations show that the most metal-rich stars ([Fe/H] $>$ 0) are, overall, the kinematically coolest and show a triangularly peaked dispersion profile at low latitudes ($b = 5^\circ$) and a flatter dispersion profile at high latitudes ($b = 10^\circ$), across longitude (see e.g. Fig. \ref{fig:rot_disp}). 

Furthermore, the shape of the dispersion profile for the low latitude value $|b|=5^\circ$ changes in the simulation. The dispersion profile for the lowest age bin shows a flatter dispersion profile at $b=5^\circ$ than the older stars and at $b=10^\circ$ it shows structure in the dispersion profile as a function of longitude. From a visual inspection, we find that this structure at $b=10^\circ$ is  due to the line of sight selection of stars that are crossing the arms of the X, and comprises the stars confined to the X-shape orbits. Lower dispersion is indicative of stars that most strongly trace out the X which are on coherent orbits.  The triangular shape that is seen in the simulation at low latitudes of $b=5^\circ$, is due to an intermediate and old stellar population (this peaked morphology is not present for the youngest stars in the second panel from the left). 

We should also point out that this simulation does not reproduce two of the MW properties:  1) the almost latitude independent dispersion of old stars as seen in the MW for stars with --0.5 $>$ [Fe/H] $>$ --1.0, and 2) the presence of a very dynamically hot population with [Fe/H] $<$ --1.0 \citep{Ness2013a}.
That the model does not reproduce the kinematics of the most metal poor stars indicates that there is some population missing from the model \citep[similar to][]{Debattista2016}.  We conclude that our model galaxy is able to qualitatively well reproduce the overall rotation and dispersion profiles seen for the MW, but again there might be a population that is missing in this division by age that matches the kinematics of the most metal poor stars. In the ARGOS survey these show a latitude independent hot dispersion and the RR Lyrae stars in the MW show negligible rotation and significantly hotter dispersion than the more metal rich stars. We will further elaborate on the  different components  building up these profiles and the differences to the observed profiles in a follow-up investigation.

\subsection{Rotation and dispersion maps}

\begin{figure}
\includegraphics[width=\columnwidth]{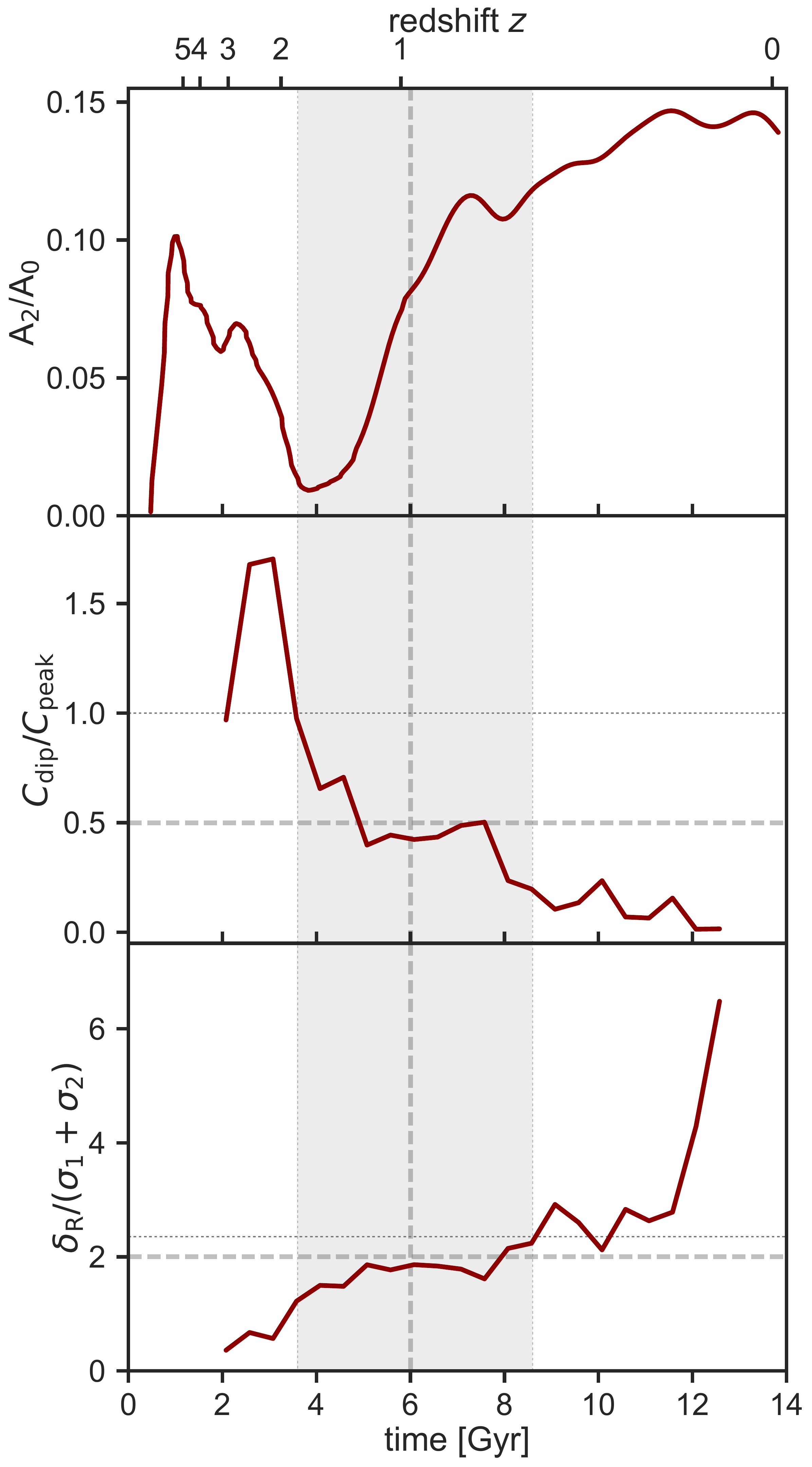}
\vspace{-.35cm}
\caption{\emph{Upper panel}: Bar strength as a function of time measured within 2 kpc from the center of the galaxy. \emph{Middle panel}:  Ratio of the dip height to  peak height of the double peaked radial distribution of stars for stars in different age bins of width $500$ Myr. \emph{Bottom panel}: Ratio of  peak separation to the sum of the widths of the two peaks of the radial distribution of stars as a function of radius for stars in different age bins of width $500$ Myr. The vertical dashed line indicates where for the first time a strong bar can be visually detected in our simulation. Faint dotted lines indicate lower and upper bounds for the point in time when the bar forms as measured by our method and the shaded area indicates the resulting uncertainty (see main text for detailed description).
}
\label{fig:bar_diagnostic}
\end{figure}

Observational surveys, particularly those that target distant stars in the bulge [e.g. APOGEE \citep{Apogee}, ARGOS \citep{Freeman2013}, Gaia-ESO \citep{Gilmore2012}, GIBS \citep{Zoccali2014}] typically adopt a pencil beam survey approach as completeness in coverage is expensive.  In Fig. \ref{fig:maps} we complement the rotation and dispersion measurements of Fig. \ref{fig:rot_disp} done in only 18 bins in $(l,b)$ by all-sky maps in $(l,b)$ of the same measurements. The left column shows the surface density maps, the middle column shows the mass-weighted rotation maps and the right panel shows the mass-weighted dispersion maps for all stars within $R=25$ kpc from the center of our simulated galaxy. Again we divide our simulation into different age bins. The upper row shows the maps for all stars, the second row shows only the youngest stars ($<2.5$ Gyr), the third row shows stars with ages $2.5<t_{\rm star}<6.0$  Gyr, the fourth row shows stars with $6.0<t_{\rm star}<10.0$  Gyr and the bottom row shows only the oldest stars with $t_{\rm star}>10.0$  Gyr.
 
From the surface density plots in the left column we see that the scale height of the disc increases for increasingly older populations while the scale length decreases \citep[see e.g.][]{Stinson2013a,Haywood2013,Marinacci2014,Bovy2016,Ma2017}. In the surface density map of all stars we see the peanut-shaped (X-shaped) structure of the bulge, and by comparing the surface density maps of different age populations we see that this structure is most prominent in the two intermediate age bins. The youngest stars are too concentrated to the mid-plane to exhibit the strong features high above the plane while the oldest population is too spherically symmetric.

In the rotation maps we see that the highest values of rotation are found close to the galaxy mid-plane and in the $l$-range of $\sim5^\circ<\vert l\vert<25^\circ$. Furthermore we see that with decreasing age of the stars the rotation velocity increases across all $(l,b)$. The old stars show a more spherically symmetric, slowly rotating configuration at all (l,b), from the bulge into the disk.

The dispersion maps are most interesting and show different structures as a function of stellar age. For all populations (including for the oldest stars) we see the  lowest dispersion values in the disk mid-plane and far away from the galactic center ($-10^\circ<l<10^\circ$), while there is a peak in velocity dispersion in the galactic center. For the younger stars (see second panel from top in Fig. 8) the velocity dispersion peak in the center itself shows a X-shaped substructure while for the older stars it is more spherically symmetric. In agreement with the findings from the rotation maps, we see that the dispersion maps show the lowest dispersion values for the young stars and increasingly higher values for older stars.

\section{Key Predictions for observables}
\label{sec:results}

Having established the overall agreement of our simulation with observations of the MW bulge, we will now use this simulation to understand the  formation scenario of the bar/bulge in the simulation -- and make predictions for upcoming surveys.  We will focus in the next subsections on the age of the bar structure and the differences between stars in the bar and the surrounding disc and in Section 4 we will investigate the formation of the bulge in detail.

\subsection{Age of the bar and X-structure as measured from the split in stellar counts}
\label{sec:x-age}

\begin{figure*}
\includegraphics[width=.33\textwidth]{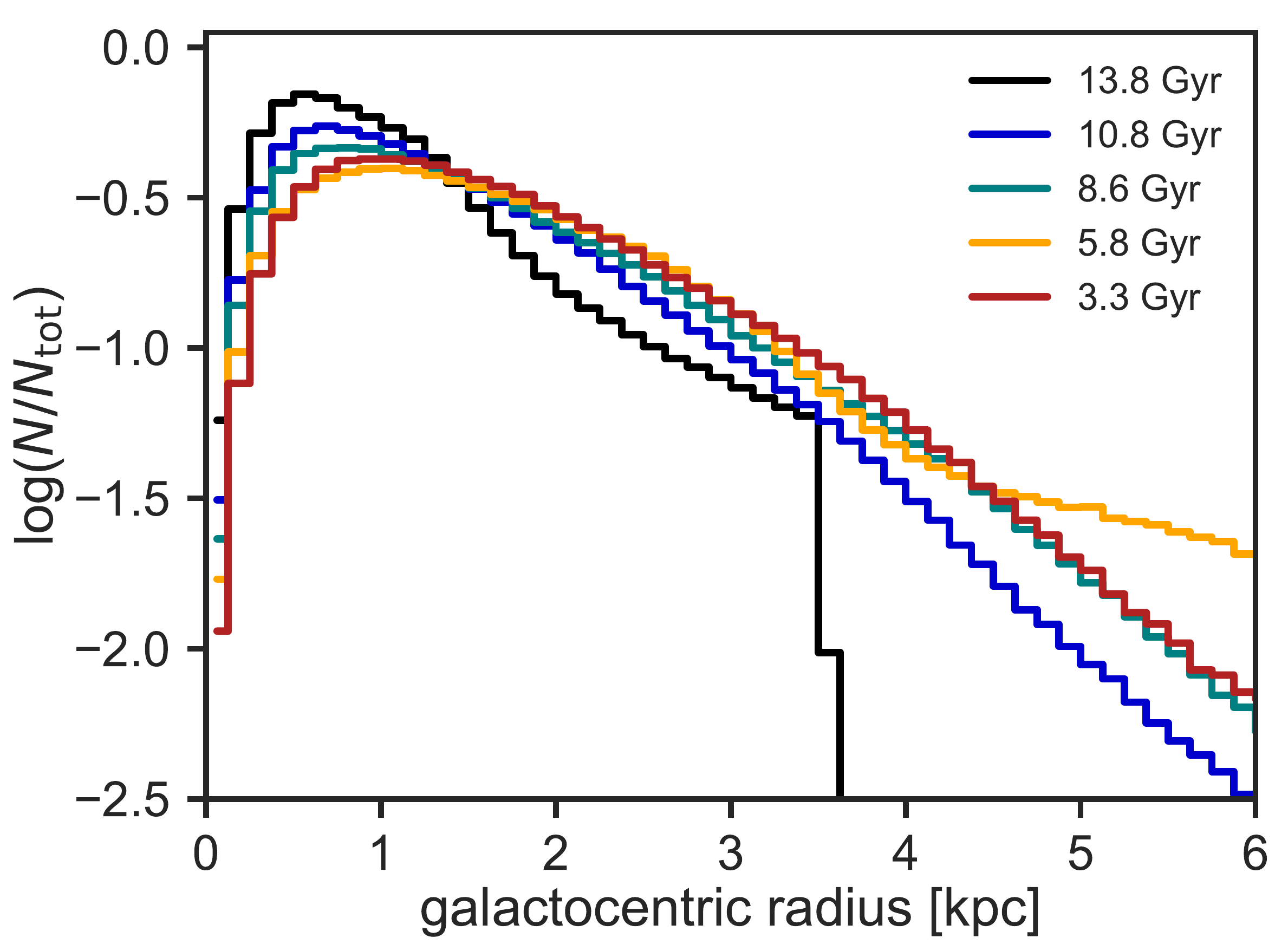}
\includegraphics[width=.33\textwidth]{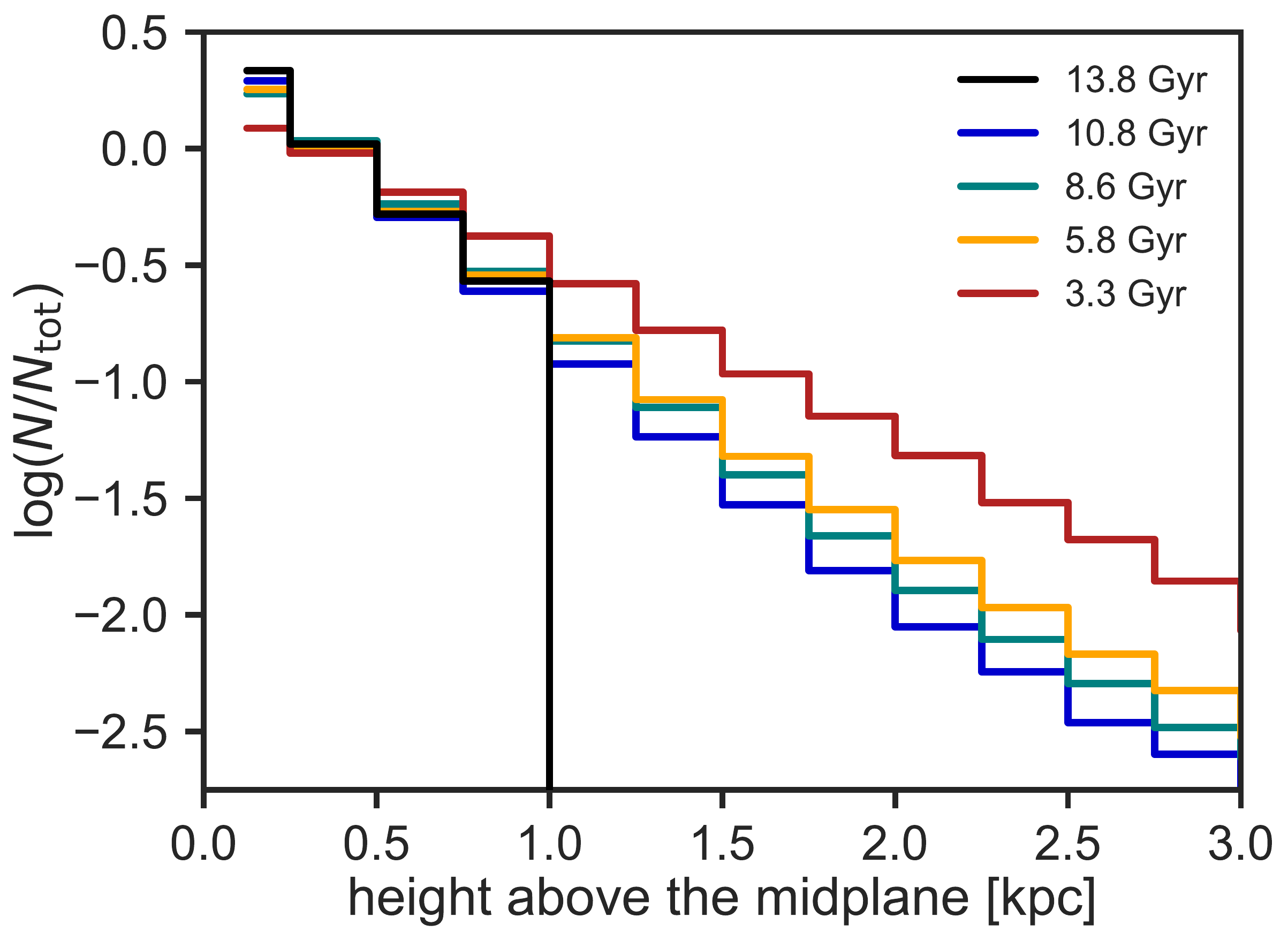}
\includegraphics[width=.33\textwidth]{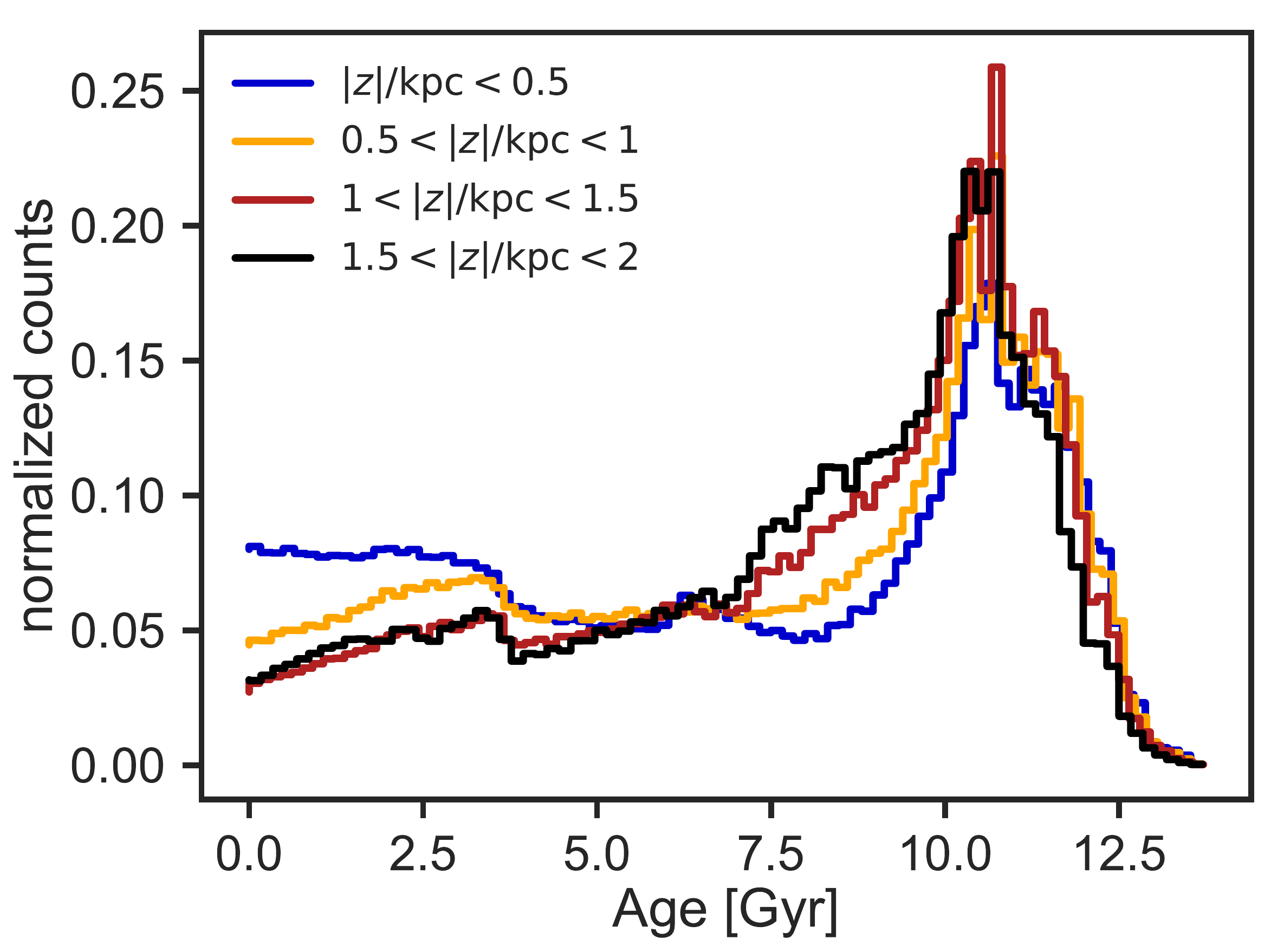}
\caption{Evolution of the radial and vertical distribution of stars in the bar at redshift $z=0$. For this figure we select stars in the redshift $z=0$ snapshot which belong to the bar and meet the following spatial selection criteria: $-3.5<x/\rm{kpc}<3.5$, $-1.25<y/\rm{kpc}<1.25$ and $-1.0<z/\rm{kpc}<1.0$.
\emph{Left panel}: Radial distribution of stars in the bar at redshift $z=0$ (black line) and  for different snapshots at earlier times in the simulation.
\emph{Middle panel}: Vertical distribution of the same selection of stars for the same selection of snapshots.
\emph{Right panel}:Age distribution of stars in the innermost 3.5 kpc from the galaxy center for different heights above the stellar mid-plane. The \emph{blue histogram} shows  the age distribution for stars  with $\vert z\vert<0.5$ kpc from the mid-plane,  the \emph{orange one} shows stars with $0.5< \vert z\vert/\rm{kpc}<1.0$ kpc,  the \emph{red histogram} shows  stars with $1.0< \vert z\vert/\rm{kpc}<1.5$ kpc and the \emph{black histogram} shows stars with $1.5 <\vert z\vert/\rm{kpc}<2.0$ kpc.
}
\label{fig:trace_stars}
\end{figure*}

In Fig. \ref{fig:bar_radius} we have seen that the bar in our simulation forms around 8 Gyr ago at redshift $z\sim 1$. This corresponds well with our results from Fig. \ref{fig:split} where we found that the split in the stellar counts is visible only for the stellar populations younger than 10 Gyr. Combining this with previous findings, that the complex dynamics of stars under the influence of the bar causes the X-shaped structure \citep[see e.g.][]{DiMatteo2016,Debattista2016,Fragkoudi2017}, we propose that the split in the stellar counts might be used to determine the formation time of the bar of the MW.  We do this by linking the X-shape distribution as a function of stellar age to the time of formation in the simulation. The underlying assumption of this method is: (a) that the effect of an evolving bar on disc stars is different for stellar populations of different (radial) velocity dispersion, and (b) there is significant evolution of the (radial) stellar velocity dispersion with cosmic time, and thus with stellar age. We present here a first idea how such a measurement could be performed. We caution however, that more simulations of barred disk galaxies and more detailed studies on the dependence of the X-shaped structure on time of formation and the specific mechanisms at play are needed to result in a robust age estimate of the bar. However, this is outside the scope of this paper and we proceed by laying out the principals of our idea:
We have seen that the peak heights and peak separation of the double-peaked distribution of stars decreases with increasing stellar ages until the double-peaked feature finally disappears for the oldest stellar bins and we are left with a single-peaked distribution. \textit{Kinematic fractionation},  as described in \citet{Debattista2016}, has differentiating effects on different stellar populations.  Given this, and that  we know the formation of the bar precisely, we might be able to calibrate the time of formation to the distribution of stars as a function of their age, and apply it to galaxies like the MW.

In the simulation the procedure is as follows: We divide the stars in the simulation into different age bins separated by 500 Myr and fit a double Gaussian to the two peaks in the stellar counts as a function of distance. From this we first determine, for every age bin, the height $C_1(t)$ and $C_2(t)$ of the two Gaussians and average them to a mean peak height $C_{\rm peak}(t)$. We define the dip height $C_{\rm dip}$ as the value of the double-Gaussian in the center at $R=8$ kpc. From the ratio of the central dip height to the average peak height $C_{\rm dip}/C_{\rm peak}$, we can determine the time when the double-peaked distribution transits into a single-peaked distribution. A similar measure can be obtained from the ratio of the peak separation $\delta_{\rm R}$ to the sum of the widths of the two Gaussians $\sigma_1 + \sigma_2$. The former ratio as a function of time is shown in the middle panel of Fig. \ref{fig:bar_diagnostic}, while the latter ratio is shown in the lower panel of Fig. \ref{fig:bar_diagnostic}. If we compare the evolution of these two curves to the evolution of the bar strength, we see that either a value of  $C_{\rm dip}/C_{\rm peak}\sim0.5$ or a value of $\delta_{\rm R}/(\sigma_1 + \sigma_2)\sim2$ marks the formation of the bar in our simulation (indicated by gray dashed lines in Fig. \ref{fig:bar_diagnostic}). This agrees well with the expectation that a value of  $C_{\rm dip}/C_{\rm peak}\sim0.5$ would correspond to a clear visual separation of two equal Gaussians with roughly a separation of $3.33\sigma$ (full width at a quarter of a maximum). And a value of $\delta_{\rm R}/(\sigma_1 + \sigma_2)\sim2$ corresponds well to the  fact that for a separation of $2.355\sigma$ (FWHM) two equal Gaussians can well be distinguished from a single peak distribution (in this case $C_{\rm dip}/C_{\rm peak}\sim1$). Clearly the agreement between the two different ways of measuring the separation of the two Gaussians is not perfect. But this is due to the fact that we do not deal with two identical Gaussians, but with asymmetric Gaussians where the Gaussian peak on the near side of the bulge is higher than the one on the far side. Needless to say that the assumption of Gaussian peaks is already a simplification. We therefore indicate in the figure the extreme case of $C_{\rm dip}/C_{\rm peak}\sim1$ and correspondingly $\delta_{\rm R}/(\sigma_1 + \sigma_2)\sim2.355$  with thin dotted lines. Taking these into account and given the fact that the bar forms around $t^{\rm bar}_{\rm form}\sim8^{+2}_{-2}$ Gyr ago we are able to calibrate the values of  $C_{\rm dip}/C_{\rm peak}$ and/or  $\delta_{\rm R}/(\sigma_1 + \sigma_2)$ to $\sim0.5$ and/or $\sim2$.  However, we caution that this method is used on only a single galaxy here. Calibration against more simulations of barred spiral galaxies would be needed in order to account for possible degeneracies between actual formation time of the bar and its effects on the stellar populations of the underlying disc.  Further calibrations would also enable a more robustly determination of the values of   $C_{\rm dip}/C_{\rm peak}$ and  $\delta_{\rm R}/(\sigma_1 + \sigma_2)$, which correspond to the formation of the bar. As we detailed above we expect the bar to influence all stars present in the inner disc. Thus the formation of the bar does not coincide with the transition from a single peak to a double peaked distribution and there can be stars of all ages be found in the bar. We do not investigate the role of diffusion over time of the stars in the bulge, where the bulge stars lose the dynamical information linked to their interaction with the bar. However, the observations which show a strong correlation between  the morphology of stars in the bulge and their kinematics as a function of [Fe/H] (and in the simulation as a function of their age) is a strong indicator that this information is preserved. This  holds promise for  the use of a metric as we propose to age date the bar's formation.   Further testing and calibrating this method on a larger set of simulations should be carried to to make a better informed estimate of the formation time of the bar in the MW.

\subsection{Where do the stars in the bar come from?}

The bar (and the boxy-peanut bulge) of the simulated galaxy formed around $8$ Gyr ago, and as such two questions arise naturally: (1) what triggers the bar formation, and (2) where do the stars that currently belong to the bar came from? We have checked visually that no merger is responsible for triggering the bar instability in this simulation. However, at redshift $z=1$ when a strong bar forms, we can identify two close encounters between satellites and the main galaxy after which the bar is established (see Fig. \ref{fig:sats} in the Appendix).  This might just be a coincidence, but investigating the connection between close encounters and bar formation is outside the scope of this paper and is left for future work \citep[see][for a detailed discussion]{Zana2017}.

To address the second question, we trace backwards in time the stellar particles of the $z=0$ bar via their unique particle IDs and analyze their spatial distributions at each timestep. The procedure is as follows: We select stars at redshift $z=0$ in the center of the galaxy in a bar-like structure. First we rotate the simulation such that the stellar disc of our simulation lies in the $x-y$ plane and the long axis of the bar coincides with the $x$-axis. We then select all the stars in the region $-3.5<x/\rm{kpc}<3.5$, $-1.25<y/\rm{kpc}<1.25$ and $-1.0<z/\rm{kpc}<1.0$.  These spatial cuts agree well with a visual identification of the bar in surface density images and also agree with the assumptions of \cite{Portail2015} for the bulge region of the MW.  

Having selected these stars, we plot in Fig. \ref{fig:trace_stars} the histograms of their radial distributions (left panel) and their vertical distance from the mid-plane of the stellar disc (middle panel) for increasingly earlier times (only for those stars already born at that given time). We see that most of the stars which are at present day in the bar (black line) were born at small galacto-centric radii (the sharp cut in the black histogram at $R\sim3.5$ kpc marks our selection of the bar at $z=0$). The histograms of the radial distribution stay peaked around $R\sim1$ kpc for all previous times shown. Only a few stars migrated inwards. As a sanity check we tracked the position of stars which have been in the bar already at redshift $z=1$ down to redshift $z=0$ and find similar results. Almost all stars already in the bar at $z=1$ stay there, only few migrating outwards. These findings suggest that stars in the bar region at the present day have a pure disc origin as already suggested by \citet{DiMatteo2016} and \citet{Fragkoudi2017}. In the left panel of Fig. \ref{fig:trace_stars} we can see that there is a contribution of stars from outside 4 kpc to the present day bar. This in combination with the findings from Fig. \ref{fig:age_dist_all} suggests that there exists a mechanism which preferentially adds younger stars (thin disc stars) to the bar but less so the inner disc \citep[see also][]{Fragkoudi2017}.

The same result is found for the  vertical height of stars above the stellar mid-plane.  Stars  at earlier times in the simulation show the same height distribution as stars at redshift $z=0$. The height distribution for the whole sample stays almost unaffected. These findings might indicate that the fractionation of of the early disc into the boxy/peanut bulge was not purely kinematical but also depends on the initial structure of the thick disc as suggested by \citet{DiMatteo2016} and \citet{Fragkoudi2017}. 
The scale height is slowly decreasing with cosmic time. About 3 Gyr after the big bang we find scale heights of almost 1 kpc. This reduces to a scale height of about 470 pc at redshift $z=0.25$, or about 11 Gyr after the big bang. These scale heights are almost twice as large as the gravitational softenings of the stellar and gaseous particles ($\sim260$ pc) in this simulation and thus the disc is at all times well resolved. Observations of  stellar discs at higher redshifts $z\gtrsim1$ \citep[e.g.][]{Elmegreen2006,Elmegreen2017} show that scale heights at these redshifts are of the order of $\sim1$ kpc. This is in good  agreement with the scale heights we find for our simulation (see also Buck et al. in prep for a resolution study of stellar disc scale heights).
We have seen in Fig. \ref{fig:maps} that younger stars show a thinner configuration and are found closer to the stellar mid-plane.  This behavior is analyzed in more detail in the right panel of Fig. \ref{fig:trace_stars} where we examine the age distribution of stars as a function of height above the stellar mid-plane at redshift $z=0$. For simplicity we have selected all stars with galacto-centric radius $r<3.5$ kpc and grouped them into 4 different bins in height ($\vert z\vert/\rm{kpc}<0.5$, $0.5< \vert z\vert/\rm{kpc}<1.0$ , $1.0< \vert z\vert/\rm{kpc}<1.5$ and $1.5 <\vert z\vert/\rm{kpc}<2.0$ ). We then plot the distribution of stellar ages for every slice of height above the mid-plane. In general, we see a double peaked distribution with one old peak around  stellar ages of $\sim9-10$ Gyr and a younger peak with stellar ages around  $\sim2$ Gyr.  Comparing the younger peak with the older peak we see that there is a slight overabundance of young stars close to the disc mid-plane (blue line) while older stars are more abundant at larger heights from the disc (red and black lines). This is in concordance with the results from Fig. \ref{fig:maps} where we have seen that young stars are concentrated close to the disc mid-plane while older stars can also be found at larger heights from the disc.  These findings are also consistent with the results from \cite{Ness2014} who studied the bulge region in an isolated simulation of galaxy formation and the results from  \cite{DiMatteo2014} who studied in detail the formation of boxy/peanut bulges by means of idealized $N$-body simulation. These authors find that all stellar populations within the outer Lindblad resonance of the bar get mapped into this structure which points towards a pure disc origin of the MW bulge. A consequence of this is that that bulge and disc populations show very similar properties.

Putting together the results from Fig. \ref{fig:trace_stars} and Fig. \ref{fig:maps} we conclude that the bar/bulge region in this simulation is formed in-situ from the disc with stars belonging to this region being locked-up there. For the whole population we do not see considerable evolution in the thickness nor in the radial component. However, we do see that there are different sub-components present in the bar/bulge region showing different spatial distributions with younger stars being found closer to the disc mid-plane in agreement with observational findings from \cite{Bensby2013} and theoretical results from \cite{Ness2014}.

\subsection{Differentiating stars in the bar from stars in the surrounding disk }
\label{sec:origin}
\begin{figure*}
\includegraphics[width=.33\textwidth]{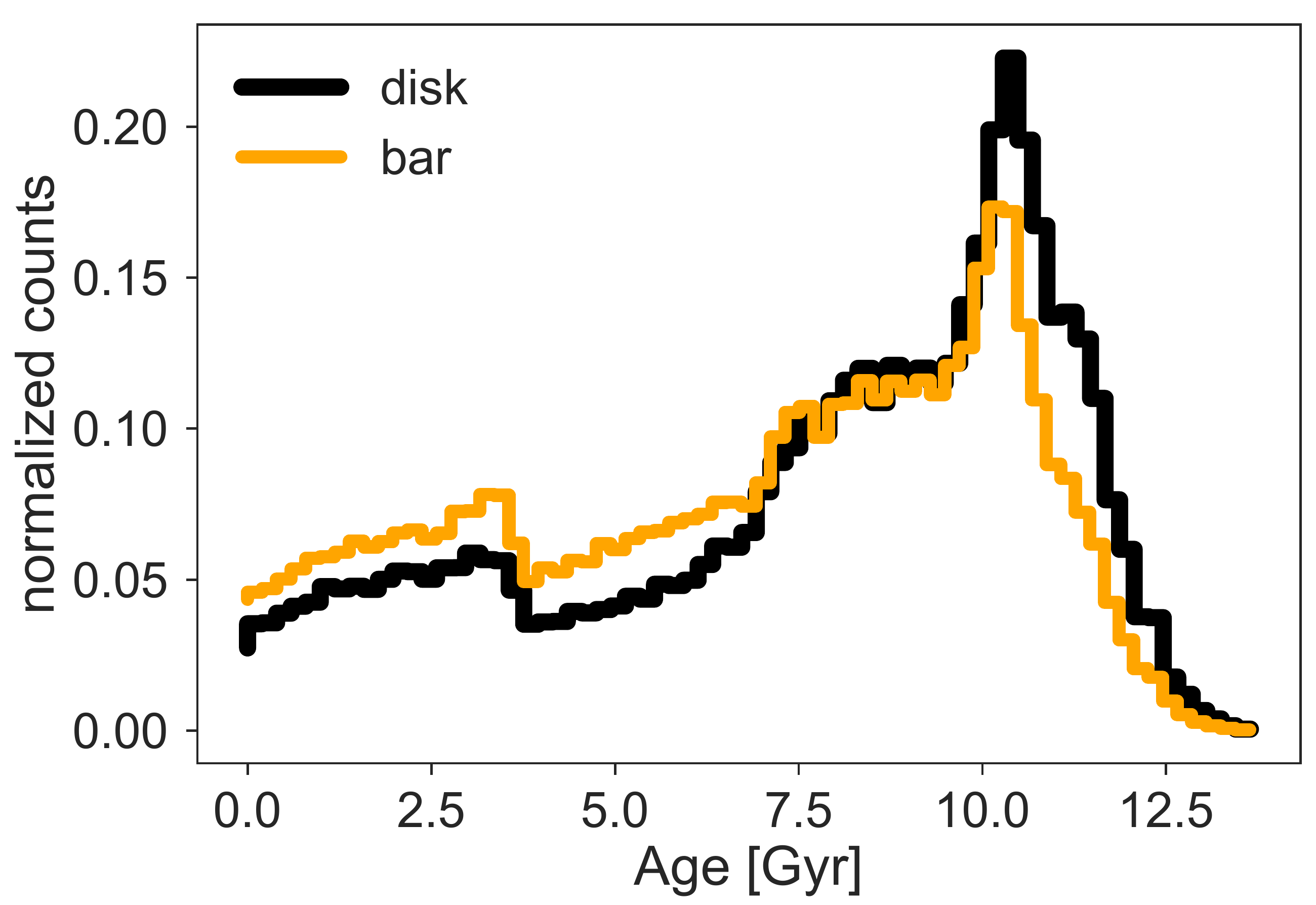}
\includegraphics[width=.33\textwidth]{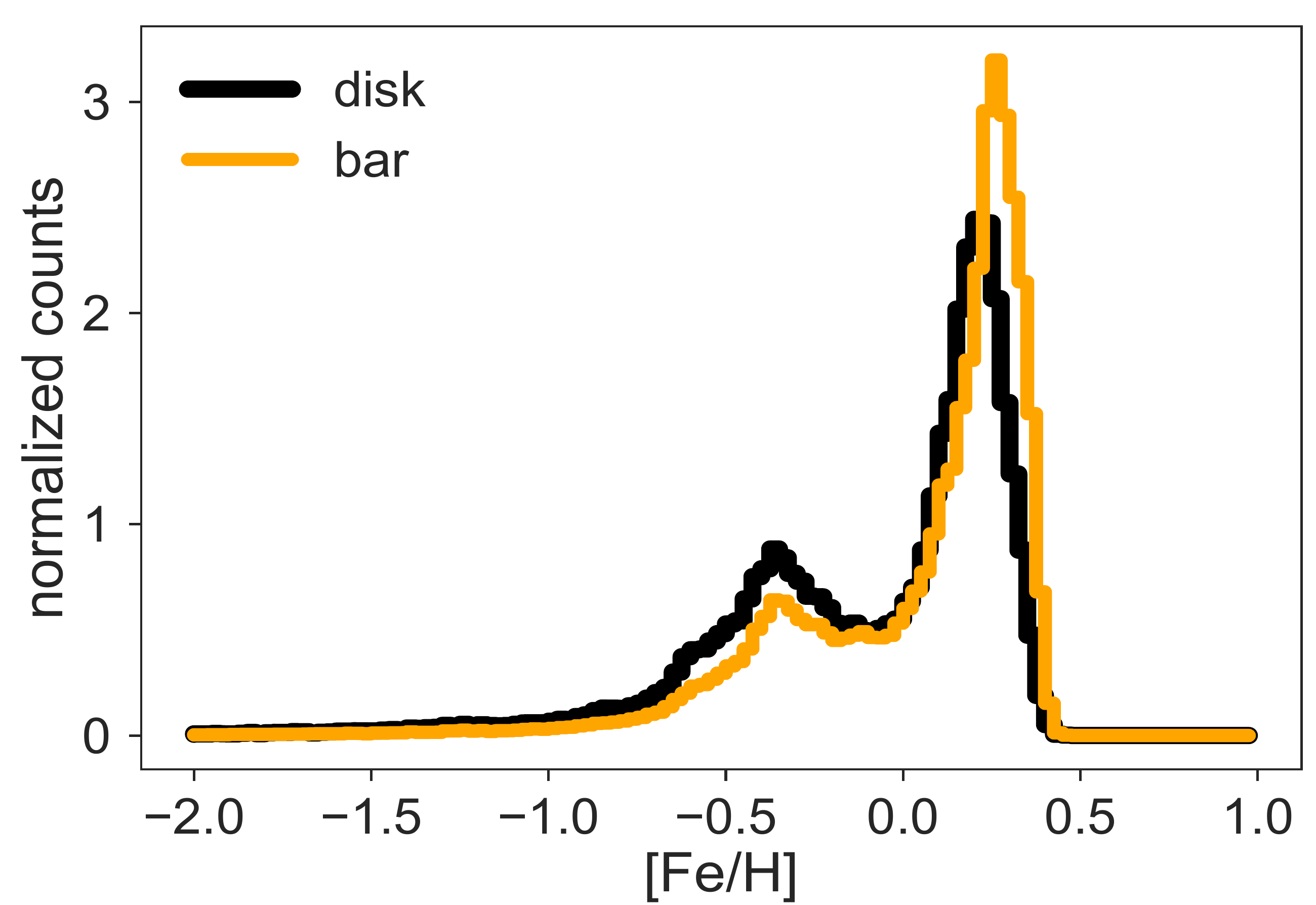}
\includegraphics[width=.33\textwidth]{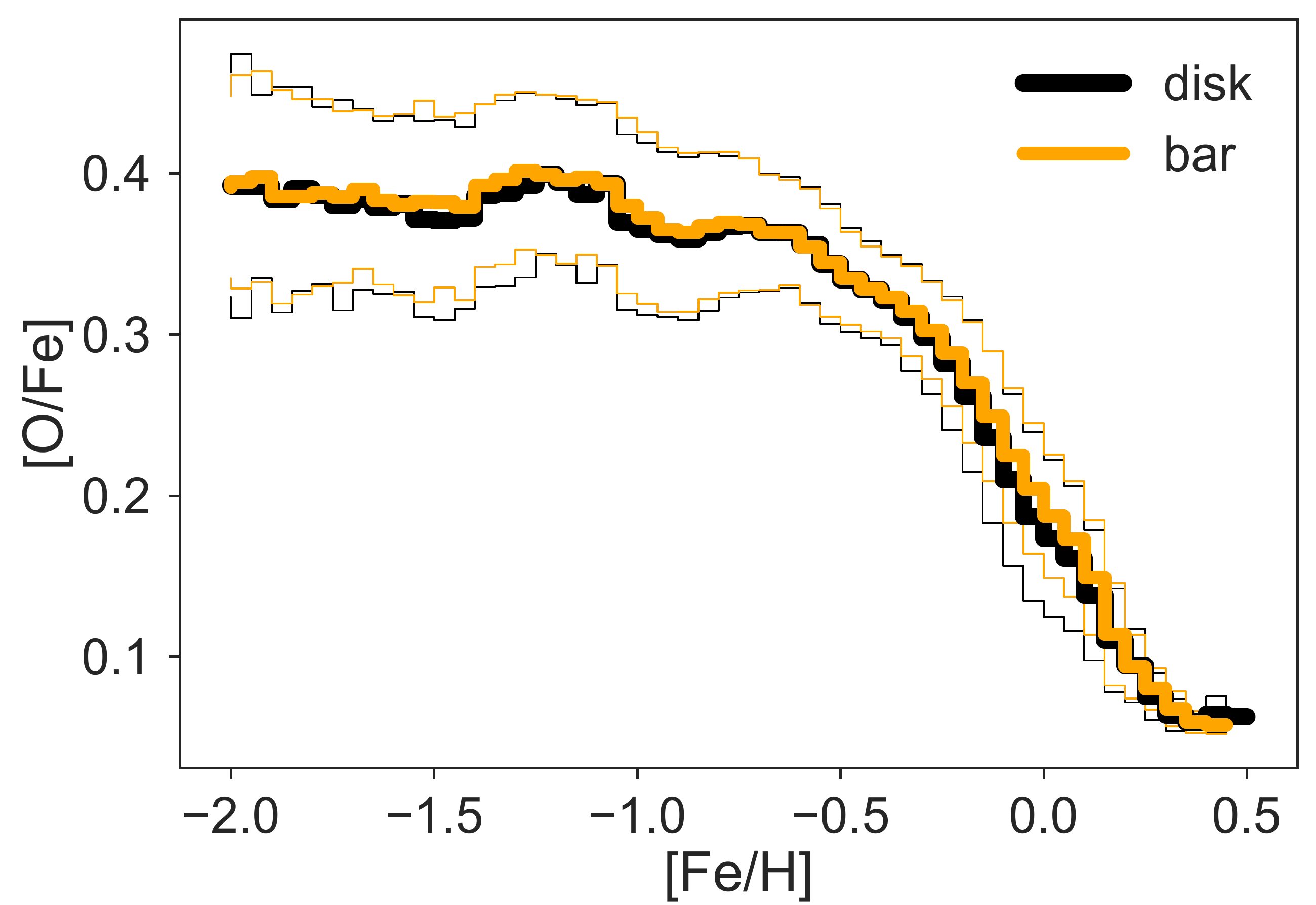}
\vspace{-.3cm}
\caption{Properties of bar (orange histogram) and "disc" stars (black histogram). The \emph{left panel} shows the age distribution function for bar and disc stars. The \emph{middle panel} shows the metallicity distribution function and the \emph{right panel} shows the oxygen abundance [O/Fe] as a proxy for $\alpha$-elements vs. metallicity.  The thin lines show the 1$\sigma$ scatter in [O/Fe]. For the other two panels the scatter is much smaller than the line thickness.}
\label{fig:bulge_met}
\end{figure*}

Given our finding that most stars in the bar were locked up in this structure since they were born, or at least shortly after their birth, we now try to answer the question if bar membership comes with a distinct imprint on this stellar population, thus enabling disk and bar to be distinguished.\footnote{That is, other than via their orbits, which are not a direct observable. (Even with Gaia, observing the bulge is problematic due to both reddening and crowing and the precision of proper motions for stars that are observed at the distance of the bulge is too low over the 5-year baseline of the mission to determine orbits for stars in much of the bulge region).} To test this, we select two samples of stars: A) The bar sample, for which we select all the stars of the bar which have a distance from the galaxy center in the range $2<R/\rm{kpc}<3.5$ and which intersect with our previous bar selection of $-3.5<x/\rm{kpc}<3.5$, $-1.25<y/\rm{kpc}<1.25$ and $-1.0<z/\rm{kpc}<1.0$   for the bar lying in the $x-y$-plane and being aligned with the $x$-axis. B) The disk sample, for which we select stars in the stellar disc with the same radial distance from the galaxy center but instead of being located in the barred structure, we select them to belong to a barred structure $90^\circ$ offset from the bar. For these two samples of stars, we plot in Fig. \ref{fig:bulge_met} the distribution of stellar ages (left panel), the distribution of stellar metallicities [Fe/H] (middle panel) and the mean value of [O/Fe] for every metallicity bin (right panel). 

The age distribution of bar stars and disc stars looks almost the same, except for a slight shift to younger ages for the bar population, maybe indicating ongoing star formation at the tips of the bar. Similar results are found for the metallicity distribution of bar and disc stars shown in the middle panel of Fig. \ref{fig:bulge_met}. Bar stars and disc stars show very similar distributions  with the bar stars offset to slightly higher metallicities. In the [O/Fe] vs. [Fe/H] plot we do not see any differences between bar and disc stars at all. The only subtle differences between bar and disc stars fit in the picture of an in-situ formation of the bar/bulge region from the disc,  as laid out previously. The slight distinctions can be explained by the different densities of stars in the bar and the surrounding disc, which differ for each population, e.g. the oldest stars do not show a barred structure but a more spherically symmetric  distribution, while for increasingly younger stars the bar is more and more pronounced. Thus, one finds in the same volume slightly more young stars in the bar region than outside it. This explains the similarity, but also the very slight differences in age and metallicity. We find in our simulation that there is continued star formation in the center and maybe even in the bar such that new, young stars are continuously added to the bar.  The near identical properties of the disk and bar populations that we find in our simulation is aligned with observational results \cite[e.g.][]{Brito2010, Bensby2017}. That the inner disk and bulge are practically indistinguishable is not an unexpected result given that the bar formed from the disk at early times. In the simulation, the differences in the overall distributions are simply a consequence of the different spatial profiles of these structures. 

\section{Summary and Conclusion}

In this study we presented a high resolution cosmological hydrodynamical simulation of a galaxy, whose bulge properties are in remarkable good agreement with MW observations. We used this simulation to study in detail the different stellar constituents of the bulge and bar region, their kinematical and chemical properties and their origin, and to make predictions for up-coming spectroscopic surveys like  MOONS \citep{C2012}, 4-MOST \citep{deJong2015},  APOGEE-2 \citep{Zasowski2017} and Sloan V \citep{Kollmeier2017}.
Our main results can be summarized as follows:

\begin{itemize}[leftmargin=*,labelindent=5pt]
\item We compare stellar counts in our simulation to the key observations of a double-peaked distribution in the line-of-sight star counts towards the Galactic Center, and find an excellent qualitative agreement between the two. All stars younger than $10$ Gyr in our simulation show a split in the line-of-sight counts with increasing peak separation for younger stars (see Fig. \ref{fig:split} ).
\item We find that in our simulation the bar leads to a fractionation of the boxy/peanut bulge. By tracing the different stellar populations of the bulge region back in time, we find that the kinematic properties of the stars are in agreement with the idea of \emph{kinematical fractionation} \citep{Debattista2016}. We also find a significant contribution from young stars to the bar which implies further dynamical effects present in the already barred galaxy.
\item The secular evolution under the influence of the bar separates initially co-spatial populations of stars into different orbit families, thus resulting in a different strength of the X-shaped structure for different stellar populations \citep[see also][for the dependence of fractionation in boxy/peanut bulge on the initial structure of the early thick disc]{DiMatteo2016,Fragkoudi2017}.
\item In Fig. \ref{fig:rot_disp} we compare the kinematics of the stars in the bulge region of our simulation with observed kinematics of the MW bulge stars taken from ARGOS, and find an excellent qualitative agreement. The shape of the rotation and dispersion profiles agree very well, although the rescaled (absolute) values for rotation and dispersion of this simulation are lower (higher) than those observed for the MW due to a higher stellar mass of the simulated galaxy. In the same figure we show the rotation and dispersion profiles for different age populations of stars in our simulation predicting what one would see if ages were available for MW stars.
\item The peak separation in the split of star counts along the line-of-sight and the age of the stellar population are correlated (Fig. \ref{fig:split}). Younger stars show larger peak separation. Furthermore, we find, in agreement with results from \cite{Debattista2016}, that the split is due to the bar separating different stellar populations. We propose that this might be used to measure the age of the bar in the MW. Using a Fourier analysis of our simulation (see Fig. \ref{fig:bar_radius}) we find an age of $\sim8$ Gyr for the bar and we calibrate a measurement involving stellar line-of-sight counts of stars in the bulge.
\item Most of the bar stars at $z=0$ were born at small radii (Fig. 10). Furthermore, our simulation suggests once stars are in the bar they are locked up there. 
\item We compare the properties of stars in the bar (with radii between 2 and 3.5 kpc) and the inner disc (same radii but $90^\circ$ offset from the bar). We find almost indistinguishable age, metallicity and oxygen abundance distributions,  (compare Fig.  \ref{fig:bulge_met}). We conclude that simple cuts in age, metallicity or oxygen abundance are not sufficient to discriminate stars that reside in the disc from those in the bar.
\end{itemize}

\acknowledgments
\section*{Acknowledgments}
We would like to thank the referee for carefully reeding our manuscript and for providing very constructive comments and suggestions which substantially helped improving the quality of the paper.
The authors like to thank Hans-Walter Rix for very fruitful discussions and exceedingly useful and inspiring comments on this work. TB and MKN are grateful to Victor Debattista for valuable suggestions and supporting discussions during the Piercing the Galactic Darkness conference. TB likes to further thank Ortwin Gerhard for his very useful comments on this topic and Dustin Lang for providing the WISE data needed to create Fig. \ref{fig:wise_comp} and for support in handling the data. TB acknowledges support from the Sonderforschungsbereich SFB 881 “The Milky Way System” (subproject A2) of the German Research Foundation (DFG).  M. Ness acknowledges funding from the European Research Council under the European Union's Seventh Framework Programme (FP 7) ERC Advanced Grant Agreement n. [321035]. AO is funded by the Deutsche Forschungsgemeinschaft (DFG, German Research Foundation) -- MO 2979/1-1. Simulations have been performed on the THEO cluster of the Max-Planck-Institut fuer Astronomie at the Rechenzentrum in Garching and the HYDRA and DRACO clusters at the Rechenzentrum in Garching. Further computations used the High Performance Computing resources at New York University Abu Dhabi.  We greatly appreciate the contributions of all these computing allocations.
This research made further use of the {\sc{pynbody}} package \citet{pynbody} to analyze the simulations and used the {\sc{python}} package {\sc{matplotlib}} \citep{matplotlib} to display all figures in this work. Data analysis for this work made intensive use of the {\sc{python}}  library  {\sc{SciPy}} \citep{scipy}, in particular {\sc{NumPy and IPython}} \citep{numpy,ipython}.
 
\bibliography{astro-ph.bib}
\newpage
\appendix

\section{SFR for different metallicity bins}

\begin{figure}
\begin{center}
\includegraphics[width=0.5\textwidth]{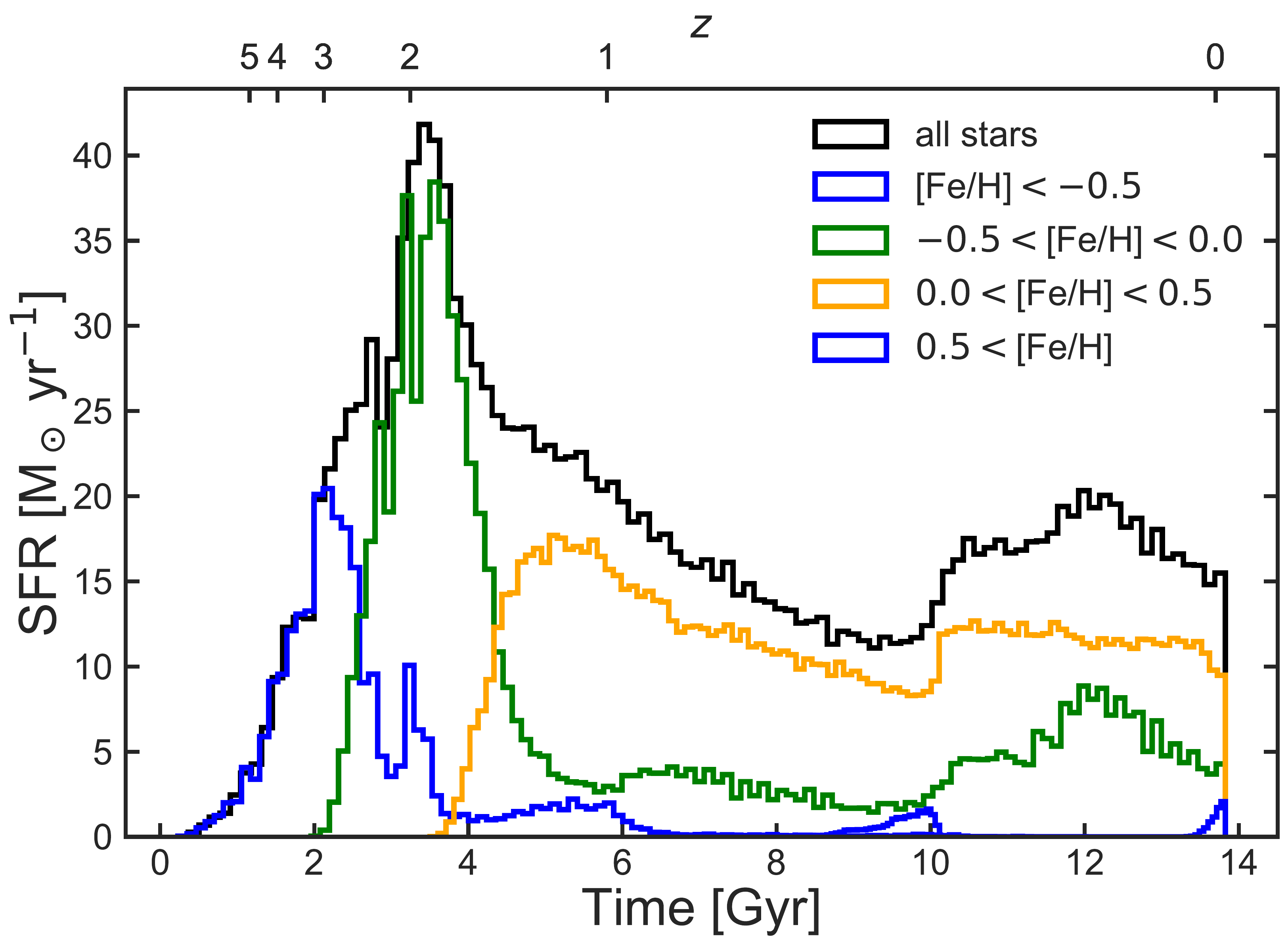}
\caption{Star formation history for stars in different metallicity bins. The black line shows the galaxy's total SF history  and colored lines shows the SF history for stellar populations of  different metallicity. Note the extended period of star formation for stars in the metallicity bin $0.0<$[Fe/H]$<0.5$.
}
\end{center}
\label{fig:SFR}
\end{figure}

\begin{figure}
\includegraphics[width=0.5\textwidth]{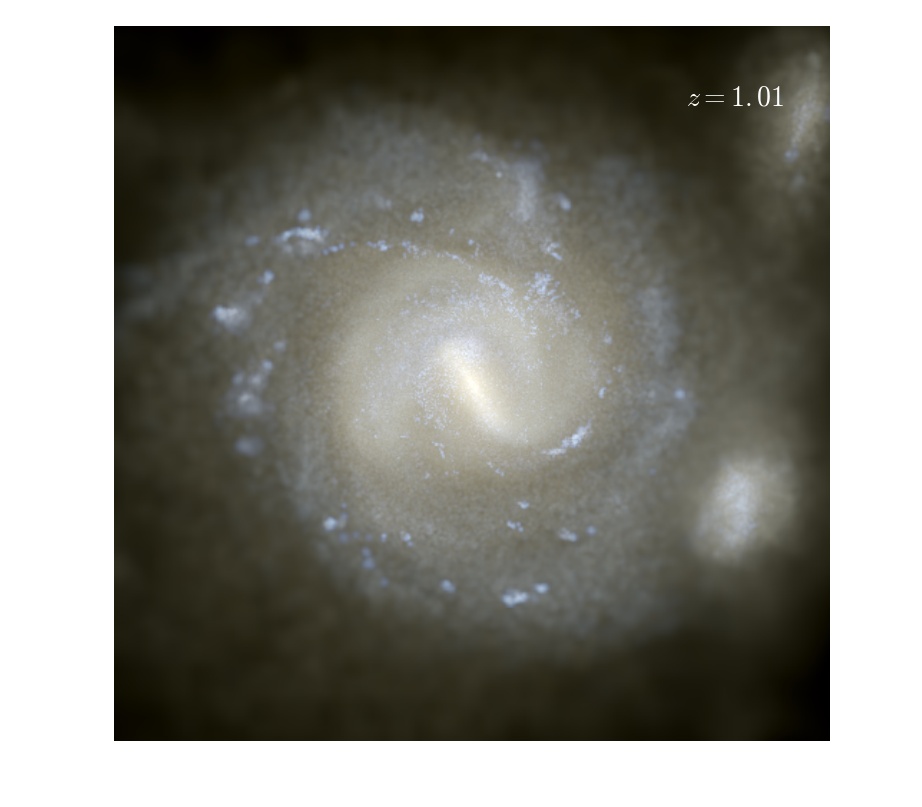}
\includegraphics[width=0.5\textwidth]{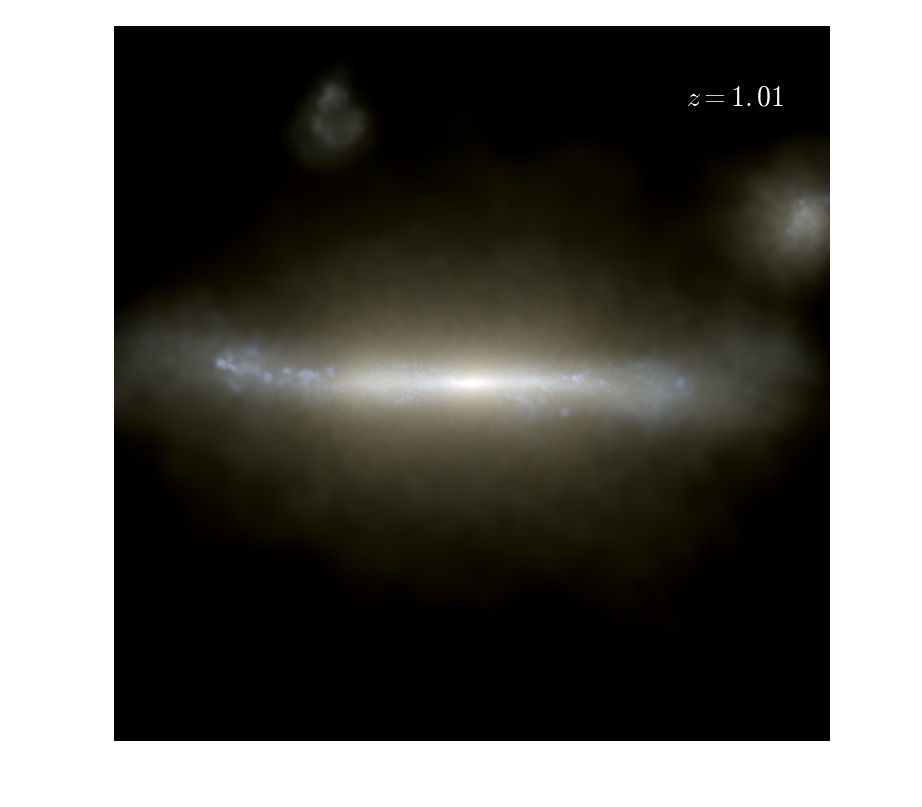}
\vspace{-.35cm}
\caption{RGB image of the galaxy at redshift $z=1$ in face-on (left panel) and edge-on (right panel) view. This figure shows the snapshot from which onwards a strong bar is visible in the simulation. Interestingly this is also the point in time at which we can observe a close encounter of two satellites with the main main galaxy. The rendering technique used is the same as for Fig. \ref{fig:rgb} in the main text.
}
\label{fig:sats}
\end{figure}

\end{document}